\documentclass[authoryear,preprint,review,12pt]{elsarticle}

\usepackage{graphics}

\usepackage{amssymb}
\usepackage{bm}
\usepackage{natbib}

\usepackage{fancyhdr}
\usepackage{latexsym}
\usepackage[abs]{overpic}
\usepackage{amsfonts}
\usepackage{amsmath}
\usepackage{mathrsfs}
\usepackage{graphicx}

\def\Tr{{\rm \,Tr\,}}
\newcommand{\pp}[1]{\phantom{#1}}

\newtheorem{theorem}{Theorem}

\newcommand{\be}{\begin{eqnarray}}
\newcommand{\ee}{\end{eqnarray}}

\journal{Ecological Modeling}

\begin{document}

\begin{frontmatter}



\title{Systems, environments, and soliton rate equations (II): Toward realistic modeling}


\author{Maciej Kuna}
\address{Katedra Rachunku Prawdopodobie\'nstwa i Biomatematyki,
Politechnika Gda\'nska, 80-952 Gda\'nsk, Poland}
\begin{abstract}
In order to solve a system of nonlinear rate equations one can try to use some soliton methods. The procedure involves three steps: (1) Find a `Lax representation' where all the kinetic variables are combined into a single matrix $\rho$, all the kinetic constants are encoded in a matrix $H$; (2) find a Darboux-Backund dressing transformation for the Lax representation $i\dot \rho=[H,f(\rho)]$, where $f$ models a time-dependent environment; (3) find a class of seed solutions $\rho=\rho[0]$ that lead, via a nontrivial chain of dressings $\rho[0]\to \rho[1]\to \rho[2]\to\dots$ to new solutions, difficult to find by other methods. The latter step is not a trivial one since a non-soliton method has to be employed to find an appropriate initial $\rho[0]$. Procedures that lead to a correct $\rho[0]$ have been discussed in the literature only for a limited class of $H$ and $f$. 
Here, we develop a formalism that works for practically any $H$, and any explicitly time-dependent $f$. As a result, we are able to find exact solutions to a system of equations describing an arbitrary number of species interacting through (auto)catalytic feedbacks, with general time dependent parameters characterizing the nonlinearity. Explicit examples involve up to 42 interacting species.
\end{abstract}


\end{frontmatter}



\section{Introduction}

The idea that formally `quantum' structures may have applications in ecology is not new, and can be traced back at least to the works of J{\o}rgensen \citep{Jorg1990,Jorg1995,JorgFath}, and (implicitly)  Ulanowicz \citep{Ulanowicz(1997),Ulanowicz(1999),Ulanowicz(2009)}. J{\o}rgensen makes an explicit reference to uncertainty principles, whereas Ulanowicz stresses the role of propensity theory \citep{Popper}. The fact that propensity is naturally related to quantum probability was intuitively felt by Popper himself, but a more recent analysis \citep{Ballentine} makes it very clear that quantum probabilities are in fact propensities. 

A theory that involves propensities and uncertainty relations cannot be formally very different from quantum mechanics, at least from a certain point of view. 
That such a possibility exists is not so surprising if one takes into account that the so-called quantum structures occur in many areas of science, and are in fact ubiquitous \citep{Aerts1986,Khrennikov(2010)}. 
If one adds that ecological models necessarily involve autocatalysis, one is almost inevitably led to the formalism of nonlinear density matrix equations \citep{ACKS}. The latter observation may be regarded as an outline of a whole research program for mathematical ecology. The present paper presents some results of this program, generalizing and extending the scope of \citep{ACKS}. The aim is to go beyond proof-of-principle and toy models, discussed in the literature so far, and develop a formalism that is flexible enough for real-life modeling.

Nonlinear density-matrix (von Neumann) equations discussed in \citep{ACKS} belong to a broad class o soliton systems.  The equations were originally discovered in the context of fundamental physics \citep{MC93,MC97} as a candidate theory for a putative nonlinear generalization of quantum mechanics. It was only later understood \citep{LC98} that the system of differential equations one arrives at is in fact a soliton one.  

A practical definition of a soliton system is the one where soliton techniques apply. Since soliton systems are known to possess a kind of universality, it is quite typical that a soliton equation discovered in one domain of science finds applications in other, completely unrelated fields. The classic example is the so-called nonlinear Schroedinger equation whose applications range from waves on deep ocean \citep{Zakharov} to optical solitons \citep{Kibler}. It was not so surprising that the same happened with soliton von Neumann equations which turned out to be equivalent to systems of coupled ordinary differential equations  similar to rate equations occurring in biological and chemical modeling \citep{Aerts et al.(2006), Aerts Czachor(2006)}, while with a slight change of interpretation their dynamics could be related to replicating two-strand quantum systems, formal analogues of  DNA \citep{Aerts Czachor(2007)}, or to $n$-level atoms in quantum optics \citep{CDSW}. In yet another reformulation, von Neumann soliton equations were found to contain as particular cases various known or new lattice systems \citep{CCU}.

The greatest advantage of soliton systems is the possibility of solving them exactly. The method, employed here and originally introduced for a simple quadratic nonlinearity in \citep{LC98},  belongs to the class of Darboux-Backlund dressing transformations \citep{DL2007}. The essence of the technique lies in finding a transformation which maps one solution $\rho(t)$ of a given equation into another solution $\rho[1](t)$ of the same equation. Not all systems of differential equations possess this property but those that do, belong to a soliton class. Once we have found the transformation it remains to find a `seed' solution $\rho=\rho[0]$ which will allow us to start the chain of transformations: $\rho[0]\to\rho[1]\to\rho[2]\to\dots$. A difficulty is that this initial step may be highly nontrivial. 

There are two problems. The first one is obvious: $\rho$ has to be found by other means. Sometimes it is easy to find a seed solution. For example, the celebrated Korteweg-de Vries soliton equation has a trivial zero solution which is nevertheless nontrivial enough to start the chain of transformations, leading to a solitary wave already after a single step \citep{Salle}. In the von Neumann case $\rho=0$ implies $\rho[1]=0$, a fact illustrating the second difficulty. Namely, an appropriate theorem  guarantees that a solution $\rho$ will generate a solution $\rho[1]$. However, the theorem does not guarantee that we will be happy with $\rho[1]$. Examples are known where, after tedious calculations, one arrives at $\rho[1](t)=\rho(t)$, or $\rho[1](t)=\rho(t-t_0)$. The art of soliton modeling is to find appropriate seed solutions by means of non-soliton methods. 

Let us illustrate the point on the simplest  yet highly nontrivial example  of a soliton von Neumann equation,
\be
i\dot\rho=[H,\rho^2]\label{1}
\ee
solved for the first time by soliton methods in \citep{LC98}. Here $\dot\rho=d\rho/dt$ and $[A,B]=AB-BA$ is the commutator. $H$ is an operator which, from the point of view of rate equations, encodes the values of possible kinetic constants \citep{ACKS}. So, we have to begin with a solution $\rho$ of (\ref{1}) and then we know, by the theorem, that some $\rho[1]$ will again satisfy 
\be
i\dot\rho[1]=[H,\rho[1]^2].\label{1'}
\ee
But how to find an initial $\rho$ if we do not know how to solve (\ref{1})?
A first try is a $\rho$ which satisfies $\rho^2=\rho$, so that $\rho(t)=e^{-iHt}\rho(0)e^{iHt}$ is a solution of (\ref{1}) since 
\be
i\dot\rho=[H,\rho^2]=[H,\rho].\label{1''}
\ee
Eq.~(\ref{1''}) is mathematically identical to the liner von Neumann equation from quantum mechanics, so we know everything about it.
However, it turns out that in such a case $\rho[1]^2=\rho[1]$ as well, and the dynamics of the new solution is effectively as {\it linear\/} as the one of $\rho$.
One of the tricks that give the right seed solution, invented in \citep{LC98}, is to  find a $\rho$ satisfying $\rho^2=a\rho+\Delta$, where $a$ is a number and $\Delta$ is an operator commuting with $H$.
Then
\be
i\dot\rho=[H,\rho^2]=a[H,\rho]+[H,\Delta]=a[H,\rho],
\ee
is effectively linear and can be easily solved. Still, an appropriate choice of $a$ and $\Delta$ guarantees that $\rho[1]$ is qualitatively different from $\rho$, and some new purely nonlinear effects occur. Note that $\Delta=F(H)$, for some function $F$, so possibilities of finding an appropriate $\rho$ may crucially depend on properties of $H$.

\begin{figure}
\centering
\includegraphics[width=\textwidth]{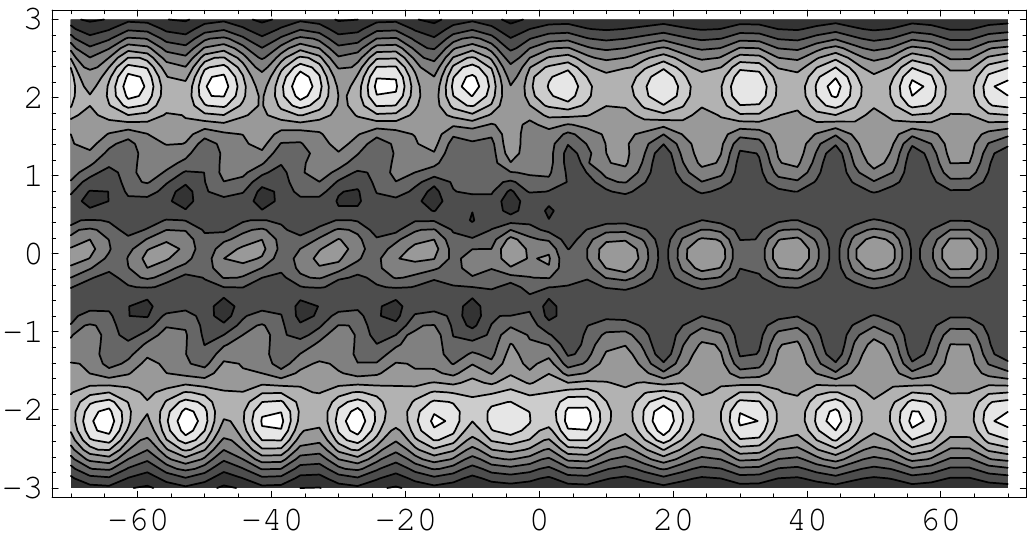}
\caption{Change of pattern as a result of self-scattering.}
\label{Fig1}
\end{figure}

The solutions $\rho[1]$ found in \citep{LC98} exhibited a new type of soliton phenomenon termed  self-scattering. The theorem on dressing transformations was further generalized in \citep{UCKL} to the general equation
\be
i\dot\rho=[H,f(\rho)]\label{f}
\ee
where $f(x)$ was basically arbitrary, and explicit solutions were found. Subsequent works showed that self-scattering may look similar to opening of a double spiral \citep{Aerts Czachor(2006)}, replication \citep{Aerts Czachor(2007)}, or morphogenesis  \citep{Aerts et al.(2003)}. Self-scattering also leads to a specific form of pattern formation, as shown in Fig.~\ref{Fig1}. 
Nonlinearity is here quadratic, $f(\rho)=\rho^2$, and $H$ is a quantum harmonic oscillator Hamiltonian. Formation of the pattern from Fig.~\ref{Fig1} is described by
\be
i\dot \rho(t,y,y')
=
\big(-\partial^2_{y}+\partial^2_{y'} +y^2-y'^2\big)\int dz \rho(t,y,z)\rho(t,z,y').\label{RD}
\ee
The pattern itself is obtained through a contour plot of 
$A(t,y)=\rho(t,y,y)$. Such a two-dimensional solution is sometimes termed a Harzian \citep{Harz}. The example of (\ref{RD}) shows that soliton von Neumann equations, when written in space-time variables, are mathematically similar to infinitely-dimensional reaction-diffusion systems 
\citep{Murray1977,Fife1979,CH93,Aerts et al.(2003)}, 
whose general form is
\be
i\dot X=\hat\omega X+\hat\omega_1 f(X)\label{r-d} 
\ee
where $\hat\omega=A\nabla^2$, and $A$ and $\hat\omega_1$ are, in general complex, matrices and $X$, $f(X)$ are vectors. Particular cases of (\ref{r-d}) are the Swift-Hohenberg, $\lambda-\omega$, and Ginzburg-Landau models  \citep{GinzburgLandau1950,HowardKopell1977,SwiftHohenberg1977,StewartsonStuart1971}.

A transition between the reaction-diffusion and rate-equation forms of the soliton system is obtained if one appropriately chooses the basis in the space of solutions. In the harmonic oscillator case the basis is given by Hermite polynomials times a Gaussian, which are eigenfunctions of $H$. The pattern from Fig.~\ref{Fig1} is obtained if at $t=0$ one starts with $\rho$ written as a combination of such Hermite-polynomial eigenfunctions (in the variable $y$).

The choice of a harmonic-oscillator $H$ is here not accidental, and is related to the structure of spectrum of $H$. Namely, it is known that harmonic oscillator is an example of a system whose energy levels are {\it equally spaced\/}: $E_1=E_0+\Delta E$, $E_2=E_1+\Delta E$, ... $E_{n+1}=E_n+\Delta E$,... with $\Delta E$ being independent of $n$. Many different Hamiltonians share this property. However, when one translates the von Neumann equation into a set of coupled rate equations, one finds that the `energy levels' play effectively the role of kinetic constants $k$, determining the dynamics of a kinetic (biological, chemical, ecological...) process. 

Still, it is very unlikely that a real-life modeling will encounter a case where kinetic constants possess this type of regularity. On the contrary, it is typical that one will deal with  basically arbitrary $k$s. Is it a problem? Yes, and a serious one: Practically all the explicit procedures of finding a seed $\rho$ one can find in the literature are based on $H$ whose spectrum is equally spaced, or at least contains an equally spaced subset of eigenvalues (my unpublished preprint \citep{Kuna2004} seems the only exception). So, the first goal of the present paper is to describe a method that works for {\it all\/} $H$ whose spectrum is discrete, and thus with {\it no restrictions\/} whatsoever on the possible values of admissible kinetic constants (although still not for the most general form of nonlinearity, see the last section).

The second goal is to allow for {\it arbitrary\/}, explicitly time dependent functions
\be
f\big(\rho(t)\big)
&=&\sum_{j=0}^\infty f_j(t)\rho(t)^j\label{f sum}
\ee
in (\ref{f}). The formulas we will discuss will be valid for {\it any\/} choice of $f_j(t)$. In practical examples, however, $\rho(t)$ is an $n\times n$ Hermitian matrix (corresponding to an ecosystem with up to $n^2$ populations existing at time $t$). For a finite $n$ the infinite sum in (\ref{f sum}) can be replaced  by a finite sum, no matter which $f$ we select (see Appendix). So, practically, {\it our formulas are valid for nonlinearities described by polynomials of any order, with arbitrary time dependent coefficients\/}. This is the second element extending our results beyond all that has been known on soliton von Neumann equations so far.
The number of independent populations encoded in an $n\times n$ matrix $\rho$ can be reduced by imposing constraints such as $\Tr \rho=1$ or the like. 

Now, before we describe all the necessary technicalities and examples, let us clarify one more point. It is known that ecosystems such as plankton may involve tens or hundreds of species \citep{Hutchinson(1961),HW99,WA 2005,AllesinaLevine2011}. There is practically no {\it non-soliton\/} technique that would allow us to find an exact solution for a system consisting of, say, 100 species  coupled by various catalytic and autocatalytic feedback loops. Here, we will give examples of explicit solutions $\rho[1](t)$ corresponding to 30 or 42 populations, interacting via (auto)catalytic feedbacks, in environments that change in time. But in order to achieve it, we have start with a seed solution $\rho(t)=\rho[0](t)$. 

The way we proceed is the following. We first write the Hamiltonian $H$ in a diagonal form (this is always possible since $H$ is Hermitian),
\be
H = \left(\begin{array}{cccc} H^{(1)}  & 0  & 0 &  \cdots\cr
0  & H^{(2)}  & 0 &  \cdots\cr
0  & 0  & H^{(3)} &  \cdots\cr
\vdots  & \vdots  & \vdots &  \ddots \cr
\end{array}\right),\nonumber
\ee
where $H^{(j)}$ are blocks of dimension 2 or 3. We then choose the seed solution in a block-diagonal form,
\be
\rho = \left(\begin{array}{cccc} \rho^{(1)}  & 0  & 0 &  \cdots\cr
0  & \rho^{(2)}  & 0 &  \cdots\cr
 0 & 0  & \rho^{(3)} &  \cdots\cr
\vdots  & \vdots  & \vdots &  \ddots \cr
\end{array}\right)\nonumber
\ee
and solve the equation in question. The solution $\rho(t)$ is typically not a very impressive one as involving an effectively linear coupling between pairs or triples of species. But recall that in the Korteweg-de Vries case the seed solution is just zero. What we are interested in is the solution $\rho[1]$  it generates, and perhaps also $\rho[2]$, $\rho[3]$ etc., since the procedure, once successfully started, can be iterated an arbitrary number of times. In our case, the seed solution is chosen in such a way that
\be
\rho[1] = \left(\begin{array}{cccc} \rho[1]^{(11)}  & \rho[1]^{(12)}  & \rho[1]^{(13)} &  \cdots\cr
\rho[1]^{(21)}  & \rho[1]^{(22)}  & \rho[1]^{(23)} &  \cdots\cr
\rho[1]^{(31)} &\rho[1]^{(32)}  & \rho[1]^{(33)} &  \cdots\cr
\vdots  & \vdots  & \vdots &  \ddots \cr
\end{array}\right)\nonumber
\ee
where {\it any\/} $2\times 2$, $3\times 3$, $2\times 3$, or $3\times 2$ block contains nonzero matrices. In standard terminology $\rho[1]$ is a single-soliton solution ($\rho[N]$ would be an $N$-soliton one). The interaction of all the species of a plankton-type community is described here through a soliton effect. We have to make sure that all $\rho[1]^{(kl)}\neq 0$, which is a nontrivial task described in detail in the following sections.

\section{The formalism}

\subsection{States}

A generic state of an ecosystem occurs only once in its history. This is basically why states represent propensities (tendencies) and not probabilities defined by the  frequency approach. (As opposed to quantum mechanics, properties of an ecosystem cannot be tested on a great number of identically prepared ecosystems.) A state is described by an operator (or just a matrix) $\rho$ satisfying: (a) Hermiticity $\rho^\dag=\rho$, (b) positivity $\rho> 0$, and (c) normalization $\Tr \rho=1$.
A Hermitian operator is positive if it does not have negative eigenvalues.  Any set of Hermitian positive operators $P_k$ (`propositions'), satisfying the resolution of unity $\sum_k P_k=I$ ($I$ denotes the identity operator) defines a set of propensities by the formula
$p_k=\Tr(P_k\rho)$ \citep{Ballentine}.  We say that two propensities $p_k=\Tr(P_k\rho)$ and $\tilde p_l=\Tr(\tilde P_l\rho)$ are complementary if the commutator $[P_k,\tilde P_l]=P_k\tilde P_l-\tilde P_lP_k=Q_{kl}$ is nonzero. The propensities are complementary since their standard deviations 
$(\Delta p)^2=\Tr (P^2\rho)-(\Tr(P\rho))^2$   satisfy the uncertainty relation
\be
\Delta p_k \Delta \tilde p_l\geq \frac{1}{2}|\Tr(Q_{kl}\rho)|.
\ee
The model is non-Kolmogorovian, i.e. does not satisfy all the axioms of probability formalized in \citep{Kolmogorov(1956)}.
One can say that complementary propensities are associated with different contexts. The set of propensities $p_k$ is maximal if its corresponding set of propositions $P_k$ sums to $I$. Complementary propensities belong to different maximal sets. Propensities belonging to different maximal sets do not have to sum to 1. It is sometimes useful to consider models where $\Tr\rho\neq 1$. The propensities are then replaced by more general kinetic variables, while Hermiticity and positivity of $\rho$ guarantee that the associated variables are nonnegative at any moment of time.

A single density matrix nonlinear equation can be treated as a very compact form of a set of nonlinear rate equations involving variables belonging to several maximal sets \citep{Aerts Czachor(2006),Aerts et al.(2006)}. In order to construct the nonnegative variables occurring in the rate equations one selects an orthogonal basis $\{ |n\rangle; \langle n|m\rangle=\delta_{nm}\}$ in the linear space which forms the domain of $\rho$. Then three families of propositions are constructed: the complete set of projectors on basis vectors, $\{P_n=|n\rangle\langle n|\}$, supplemented by two sets of projectors, $\{P_{jk}=|jk\rangle\langle jk|\}$ and $\{P'_{jk}=|jk'\rangle\langle jk'|\}$, constructed from linear combinations of the basis vectors
\be
|jk\rangle &=& \frac{1}{\sqrt{2}}\big(|j\rangle +|k\rangle\big),\\
|jk'\rangle &=& \frac{1}{\sqrt{2}}\big(|j\rangle -i|k\rangle\big), 
\ee
for $j\neq k$. Decomposing matrix elements of $\rho$ into real and imaginary parts, $\rho_{nm} = \langle n|\rho|m\rangle = x_{nm} +iy_{nm}$,  we obtain three families of nonnegative variables,
\be
p_n &=& {\rm Tr} P_n \rho = x_{nn}\cr
p_{jk} &=& {\rm Tr} P_{jk} \rho = x_{jk} + \frac{1}{2}(p_j + p_k )\cr
p'_{jk} &=& {\rm Tr} P'_{jk} \rho = y_{jk} + \frac{1}{2}(p_j + p_k )
\ee
It is interesting that replicator equations, a typical tool in evolutionary game theory \citep{MS,HS,FS}, can be written in a density matrix nonlinear von Neumann form, but with probabilities  interpretable as $p_n$ \citep{Pryk} ($p_{jk}$ and $p'_{jk}$ have no clear interpretation: $x_{jk}=\sqrt{p_jp_k}$, $y_{jk}=0$).
If $\rho$ is a positive but non-normalized solution ($\rho>0$, $\Tr\rho\neq 1$) then the variables are nonnegative, but cannot be treated as probabilities or propensities. 
Of particular interest is the case of a dynamics which does not preserve positivity of $\rho(t)$ for all $t$. In such a case, starting with a positive initial condition $\rho(0)>0$, we will find dynamic variables that are nonnegative only for certain finite amounts of time. This type of solution could be interpreted as a system where certain species disappear or appear after some time. Unfortunately, this very interesting possibility is beyond the scope of  the present paper.

\subsection{Hierarchic organization of environments}

Environments are organized hierarchically \citep{Allen}.
Subsystems are coupled to environments non-symmetrically \citep{ACKS}. The mathematical model is built on the basis of a hierarchically coupled set of rate equations,
\be
\dot\rho_1 &=& F_1(\rho_1),\\
\dot\rho_2 &=& F_2(\rho_1,\rho_2),\\
\dot\rho_3 &=& F_3(\rho_1,\rho_2,\rho_3),\\
&\vdots&
\ee
The system described by $\rho_1$ plays a role of environment for the remaining subsystems. The one described by $\rho_2$ is the environment for $\rho_3$, $\rho_4$, and so on. The collection of rate equations has to be solved in a hierarchical way. One begins with $\rho_1$ since the associated differential equation is closed. Once one finds a given $\rho_1(t)=r(t)$, one switches to
\be
\dot\rho_2 &=& F_2(r,\rho_2).
\ee
At each level of the hierarchy (perhaps with the exception of $\rho_1$), one has to solve a system of coupled nonlinear rate equations with time dependent coefficients. The formalism described in the present paper assumes that the time-dependent coefficients are arbitrary. All the examples discussed below can be understood as corresponding to an $n$-th level of the hierarchy.

\subsection{Dressing transformation from a Zakharov-Shabat problem}

We shall consider a Zakharov-Shabat (ZS)  problem \citep{Salle} with linear operators $A_{\mu}$ and $B$ acting on some Hilbert space ${\cal H}$:
\be
i| \dot\phi_{\mu}\rangle = (A_{\mu} +B)| \phi_{\mu}\rangle\label{zs1}.
\ee
$\mu$ is a complex parameter and $| \phi_{\mu}\rangle$ a vector in Dirac notation (a Dirac `ket'; in practical calculations typically represented by a 1-column matrix).
Additionally, in order to introduce a dressing transformation, we need a conjugate equation with an independent parameter $\nu$:
\be
-i\langle \dot\psi_{\nu}| = \langle \psi_{\nu}|(A_{\nu} +B)\label{zs2}.
\ee
$\langle \psi_{\nu}|$ is a dual vector (a Dirac `bra'; typically represented by a 1-row matrix). 
The main tool of dressing transformations is the operator
\be
P=\frac{| \phi_{\mu}\rangle \langle \psi_{\nu}|}{\langle \psi_{\nu}| \phi_{\mu}\rangle}.
\ee
From (\ref{zs1}), (\ref{zs2}) we prove that $P$ satisfies the nonlinear equation
\be
i\dot P &=& (A_{\mu} +B)P -P(A_{\nu} +B) - P(A_{\mu} +B)P +P(A_{\nu} +B)P \\
&=& (1-P)A_{\mu}P -PA_{\nu}(1-P) + BP - PB.\label{nme}
\ee
Now we take a third ZS problem
\be
-i\langle \dot\varphi_{\lambda}| &=& \langle \varphi_{\lambda}|(A_{\lambda} +B)\label{zs},
\ee
with yet another parameter $\lambda$. We assume that our dressing transformation can be constructed by linear operators $S$ and $T$ of the form
\be
\langle \varphi_{\lambda}[1]| &=& \langle \varphi_{\lambda}|S = \langle \varphi_{\lambda}|(1 + aP)\\
A_{\lambda}[1] &=& TA_{\lambda}T^{-1} = (1 + bP)A_{\lambda}(1 + \hat{b}P)
\ee
Since $P^2=P$ one finds $S^{-1} =I + \hat{a}P$ for $\hat{a} = \frac{-a}{1+a}$ and analogously $\hat{b}= \frac{-b}{1+b}$.
We demand
\be
-i\langle \dot\varphi_{\lambda}[1]| &=& \langle \varphi_{\lambda}[1]|(A_{\lambda}[1] +B)\label{zs},
\ee
and check how it restricts the form of  $S$ and $T$:
\be
-i\langle \dot\varphi_{\lambda}[1]| &=& \langle \varphi_{\lambda}|(A_{\lambda} +B)(1 + aP) \\
&\pp=&- \langle \varphi_{\lambda}|a\Big((1-P)A_{\mu}P -PA_{\nu}(1-P) + BP - PB\Big)\\
&=&\langle \varphi_{\lambda}[1]|\Big(A_{\lambda} + B + P(\hat{a} A_{\lambda} - \hat{a} A_{\nu})
 + (a A_{\lambda} - a A_{\mu})P\nonumber\\
  &\pp=&  +P(\hat{a}a A_{\lambda} +  a A_{\mu} + \hat{a} A_{\nu}  ) P \Big)
 \ee
 and
 \be
 \langle \varphi_{\lambda}[1]|(A_{\lambda}[1] +B)&=&\langle \varphi_{\lambda}[1]|\Big((1 + bP)A_{\lambda}(1 + \hat{b}P) +B\Big)\\
 &=&\langle \varphi_{\lambda}[1]|\Big(A_{\lambda} +B + P b A_{\lambda} + \hat{b}A_{\lambda}P + P\hat{b}b A_{\lambda}P\Big)
\ee
leading to the following three conditions:
\be
1)&& A_{\lambda}  = \frac{\hat{a}}{\hat{a} -b} A_{\nu}, \\
2)&& A_{\lambda}  = \frac{a}{a -\hat{b}} A_{\mu},\\
3)&&\hat{a}a A_{\lambda} +  a A_{\mu} + \hat{a} A_{\nu} = \hat{b}b A_{\lambda}.
\ee
It is easy to see that first two conditions imply the third. They also give  a relation between $A_{\mu}$ and $A_{\nu}$:
\be
\frac{1}{1+b} A_{\nu}&=&   A_{\mu}
\ee
A change of a parameter just multiplies the operator by a number, implying
$ A_{\mu} = X(\mu) A$, $ A_{\nu} = Y(\nu) A$ and $ A_{\lambda} = Z(\lambda) A$, for some operator $A$ and three functions of the parameters. So,
\be
\frac{1}{1+b} Y(\nu)&=&   X(\mu),\\
b&=&\frac{Y(\nu)}{ X(\mu)}-1.
\ee
From the second condition we find
\be
 A_{\lambda}  &=& \frac{a}{a -\hat{b}} A_{\mu},\\
 Z(\lambda)  &=& \frac{a}{a -\hat{b}} X(\mu),\\
a &=& -\frac{(Y(\nu)-X(\mu)) Z(\lambda)}{Y(\nu)(Z(\lambda)-  X(\mu))}.
\ee
Following \citep{UCKL} we set $  X(\mu)= \frac{1}{\mu}$, $ Y(\nu) = \frac{1}{\nu}$ and $  Z(\lambda)  = \frac{1}{\lambda}$, which finally gives
\be
b &=&\frac{Y(\nu)}{ X(\mu)}-1 = \frac{\mu}{\nu}-1= \frac{\mu - \nu}{\nu},\\
a &=& -\frac{(Y(\nu)-X(\mu)) Z(\lambda)}{Y(\nu)(Z(\lambda)-  X(\mu))},\\
&=& \frac{\nu-\mu }{\mu - \lambda}.
\ee
The dressing transformation has been found:
\be
S &=& I+ aP=I+\frac{\nu-\mu }{\mu - \lambda}P,\\
T &=& I+bP=I+\frac{\mu - \nu}{\nu}P.
\ee
One should be aware that various generalizations of the above procedure are known, including greater numbers of parameters, operator solutions of ZS-type problems, time-dependent parameters etc. \citep{Salle,Ci,Ci1,DL2007}. Of particular interest is the generalization proposed in \citep{Ci1} since it seems to be applicable to soliton von Neumann  systems with dissipation, a problem which only briefly mentioned in \citep{ACKS}, and which is also beyond the scope of the present paper.

\subsection{Lax pair}

Consider the pair of ZS-type problems
\be
z_{\mu} | \phi_{\mu}\rangle &=& \left( \frac{1}{\mu}\rho -H\right)| \phi_{\mu}\rangle\label{zs3}\\
i| \dot\phi_{\mu}\rangle &=& \frac{1}{\mu}A| \phi_{\mu}\rangle\label{zs4}
\ee
where $z_\mu$ is a $t$-independent complex number. Differentiating (\ref{zs3}) over $t$ and multiplying (\ref{zs4}) by $-iz_\mu$ we obtain two equations with identical left-hand sides. Compatibility of the right-hand sides is equivalent to
\be
i\dot\rho &=& [H,A],\label{nv1}\\
 \left[ \rho , A\right] &=& 0. \label{com}
\ee
The second condition means that $A$  is an arbitrary function of $\rho$, i.e. $A=f(\rho)$. Therefore the first condition is a nonlinear equation with respect to $\rho$. This is our nonlinear von Neumann equation
\be
i\dot\rho &=& [H,f(\rho)].\label{frho}
\ee
We say that (\ref{zs3})--(\ref{zs4}) is a Lax pair for (\ref{frho}), whereas (\ref{frho}) is a Lax representation of a system of rate equations. This Lax pair was found in \citep{UCKL} and later generalized to yet more complicated von Neumann type equations in \citep{UC,CCU}.

Now, let us apply the {\it same\/} dressing transformation to both equations of the Lax pair (\ref{zs3})--(\ref{zs4}).
Because we use two linear ZS problems of the  same shape they can be transformed by the same dressing transformation and the compatibility conditions for the transformed Lax pair are the same. Just replace $\rho$ and $A$ by $\rho [1] = T\rho T^{-1} = (1 + \frac{\mu - \nu}{\nu}P)\rho (1 + \frac{\nu -\mu}{\mu}P)$ and $A[1]= TAT^{-1}$. In this way the dressing transformation gives a connection between solutions of the equation of (\ref{nv1}), expressed by

\begin{theorem}
Assume $| \phi_{\mu}\rangle$  is a solution of (\ref{zs3}) and (\ref{zs4}) and $\langle \psi_{\nu}|$ is a solution of
\be
z_{\nu} \langle \psi_{\nu}| &=& \langle \psi_{\nu}|\left( \frac{1}{\nu}\rho -H\right)\\
-i\langle \dot\psi_{\nu}| &=& \langle \psi_{\nu}| \frac{1}{\nu}A.
\ee
Let $\rho [1] = T\rho T^{-1} = (1 + \frac{\mu - \nu}{\nu}P)\rho (1 + \frac{\nu -\mu}{\mu}P)$ and $A[1]= TAT^{-1}$ with
\be
P=\frac{| \phi_{\mu}\rangle \langle \psi_{\nu}|}{\langle \psi_{\nu}| \phi_{\mu}\rangle}.
\ee
In this case if $\rho$ and $A$ fulfill (\ref{nv1}) and (\ref{com}), then $\rho [1]$ and $A[1]$ satisfy (\ref{nv1}) and (\ref{com}) as well.

\end{theorem}
To prove the theorem we employ (\ref{nme}) for both ZS problems, and find
\be
[ H, P] &=&\frac{\mu - \nu}{\mu \nu}P \rho P + \frac{1}{\mu}\rho P  - \frac{1}{\nu}P \rho, \label{me1}\\
i\dot P &=& \frac{\mu - \nu}{\mu \nu}P AP + \frac{1}{\mu}AP - \frac{1}{\nu}PA. \label{me2}
\ee
Inserting (\ref{me1})--(\ref{me2}) into (\ref{nv1})-- (\ref{com}) we obtain
\be
i\dot\rho[1] &=& \big[H,A[1]\big],\nonumber\\
 \big[ \rho[1] , A[1]\big] &=& 0. \nonumber
\ee
which ends the proof. 

One similarly proves
\begin{theorem}
Assume that $\rho$ and $A$ satisfy
\be
i\dot\rho &=& [H, A], \nonumber\\
{[A, \rho]}&=&0 \nonumber
\ee
and there exists  $R$ which, for some nonzero numbers $\alpha$, $\beta$ and $\gamma$,  satisfies
\be
i\dot R &=&  \alpha R A R + \beta A R + \gamma R A,\\
{[ H, R]} &=& \alpha R \rho R + \beta \rho R  + \gamma R \rho.
\ee
Then
\be
\rho [1] &=&(1+ \frac{\alpha}{\beta} R) \rho(1 + \frac{\alpha}{\gamma} R),\nonumber\\
A [1] &=&(1+ \frac{\alpha}{\beta} R) A (1 + \frac{\alpha}{\gamma} R) \nonumber
\ee
fulfill
\be
i\dot\rho[1] &=& \big[H, A[1]\big] \nonumber,\\
{\big[A[1], \rho[1]\big]}&=&0 \nonumber.
\ee
\end{theorem}

All the examples discussed below are based on Theorem~1. Setting $\nu = \overline{\mu}$ one obtains a unitary $T$, so that the dressing transformation preserves  Hermiticity and positivity of $\rho[1]$. 

\section{Constructing seed solutions}

This is the core part of the work. We will show how to construct seed solutions for all $H$, any time-dependent $f$, and arbitrary numbers of interacting species. In order to eliminate typing errors, all the explicit examples discussed later on in the paper were cross-checked in Wolfram Mathematica.

\subsection{Building seed solutions with $2\times 2$ blocks}

Assume the Hamiltonian has discrete spectrum, $H = \sum_nh_n Q_n$, where $Q_n$ and $h_n$ are spectral projectors and
eigenvalues of $H$, respectively. We start from an even-dimensional case. Now choose two spectral projectors $Q_{k_1}$ and $Q_{k_2}$, related to different eigenvalues, generating a projector ${\bf R}^{(k)} = Q_{k_1} + Q_{k_2} $ on a two-dimensional invariant subspace ${\cal H}^{(k)}$. We are interested in solutions
$$i\dot\rho^{(k)} = [H^{(k)} , f(\rho^{(k)} ) ]\label{vnk}$$
for $\rho^{(k)} = {\bf R}^{(k)} \rho
{\bf R}^{(k)}$ and $H^{(k)} = {\bf R}^{(k)} H {\bf R}^{(k)}$.

Here, a notational digression. Whenever we perform explicit matrix calculations, we implicitly identify
\be
H^{(k)} = {\bf R}^{(k)} H {\bf R}^{(k)} \equiv \left(\begin{array}{cc} h_{1}^{(k)}  & 0 \cr
0  &   h_{2}^{(k)} \cr
\end{array}\right)
\equiv
\left(\begin{array}{cccc} \ddots  & 0  & 0 &  \cdots\cr
0  & h_{1}^{(k)}  & 0 &  \cdots\cr
0  & 0  & h_{2}^{(k)} &  \cdots\cr
\vdots  & \vdots  & \vdots &  \ddots \cr
\end{array}\right)\nonumber.
\ee
The same notation applies to $\rho^{(k)}$, $|\phi^{(k)}\rangle$, $|\varphi^{(k)}\rangle$, etc. So, although, say $|\varphi^{(1)}\rangle$ and $|\varphi^{(2)}\rangle$ are, for calculational purposes, treated as 2-dimensional column matrices, their sum $|\varphi^{(1)}\rangle+|\varphi^{(2)}\rangle$ is understood as a 4-dimensional column matrix. A mathematical purist would therefore rather write the sum as a direct sum. The convention we employ will not lead to ambiguities.

In a two-dimensional space, all functions of the operator $\rho^{(k)}$ (which is not proportional to the identity operator, so has two different eigenvalues $\lambda^{(k)}_1$ and  $\lambda^{(k)}_2$) are equal to a linear function of $\rho^{(k)}$ (see Appendix),
$$f(\rho^{(k)}) = \theta_1^{(k)} \rho^{(k)} +{\theta_0}^{(k)}{\bf R}^{(k)},$$
where   $\theta_1^{(k)} = \frac{f(\lambda^{(k)}_1 )
- f(\lambda^{(k)}_2 )}{\lambda^{(k)}_1 - \lambda^{(k)}_2 } $ and ${\theta_0}^{(k)}=f(\lambda^{(k)}_2 ) - \lambda^{(k)}_2 \theta_1^{(k)}$.
Note that although $\lambda_j^{(k)}$ are time independent, the parameters $\theta_j^{(k)}$ do depend on time if $f$ itself is time dependent. 
Therefore,  any such solution of (\ref{vnk}) has the following form:
$${\rho^{(k)} }(t) = U^{(k)} {\rho^{(k)} }(0) {U^{(k)}}^\dag = e^{-i\int_0^t \theta_1^{(k)} d\tau H^{(k)} } {\rho^{(k)} }(0) e^{i\int_0^t \theta_1^{(k)} d\tau H^{(k)} },$$
and also
$$f({\rho^{(k)} }(t)) = U^{(k)} f( {\rho^{(k)} }(0)) {U^{(k)}}^\dag = e^{-i\int_0^t \theta_1^{(k)} d\tau H^{(k)} }f( {\rho^{(k)} }(0)) e^{i\int_0^t \theta_1^{(k)} d\tau H^{(k)} }.$$
Let us stress again that $f( {\rho^{(k)} }(0))$ is time-dependent because $f$ is a polynomial in ${\rho^{(k)} }(0)$, but with coefficients which depend on time. 
Therefore, for this case
$$f({\rho^{(k)} }(t)) =e^{-i\int_0^t \theta_1^{(k)} d\tau H^{(k)} }[ \theta_1^{(k)}(t) \rho^{(k)}(0) +{\theta_0}^{(k)}(t){\bf R}^{(k)}] e^{i\int_0^t \theta_1^{(k)} d\tau H^{(k)} }.$$
It is convenient to rewrite the  the Lax pair on the subspace ${\cal H}^{(k)}={\bf R}^{(k)}{\cal H}$. First we use  $|\phi^{(k)} \rangle = {U^{(k)}}^\dag |\varphi^{(k)} \rangle
 = e^{i\int_0^t \theta_1^{(k)} d\tau H^{(k)} }|\varphi^{(k)} \rangle$, so that
\begin{eqnarray}
z_{\mu }|\varphi^{(k)}(t) \rangle &=& \left(\frac{1}{\mu }\rho^{(k)}(t) -  H^{(k)} \right)|\varphi^{(k)}(t) \rangle\cr
i|\dot\varphi^{(k)}(t) \rangle &=& \frac{1}{\mu }f(\rho^{(k)}(t) )|\varphi^{(k)}(t) \rangle ,
\end{eqnarray}
turns into
\begin{eqnarray}
z_{\mu }|\phi^{(k)}(t) \rangle &=& \left(\frac{1}{\mu }{\rho^{(k)} }(0)-  H^{(k)} \right)|\phi^{(k)} (t)\rangle\cr
i|\dot\phi^{(k)}(t)\rangle &=& \left( \frac{1}{\mu }f({\rho^{(k)} }(0)) - \theta_1^{(k)} H^{(k)} \right)
|\phi^{(k)} (t)\rangle\cr
&=& \left(\theta_1^{(k)}\left[ \frac{1}{\mu }{\rho^{(k)} }(0)  -   H^{(k)}\right] + \frac{1}{\mu }  {\theta_0}^{(k)}   {\bf R^{(k)}} \right)
|\phi^{(k)} (t)\rangle\cr
&=& \left(z_{\mu }\theta_1^{(k)} + \frac{{\theta_0}^{(k)}}{\mu }\right)
|\phi^{(k)} (t)\rangle.
\end{eqnarray}
Taking this into account, we obtain
\begin{eqnarray}
z_{\mu }|\phi^{(k)} (0)\rangle &=& \left(\frac{1}{\mu }{\rho^{(k)} }(0)-  H^{(k)} \right)|\phi^{(k)} (0)\rangle\label{st}\\
|\varphi^{(k)} (t)\rangle
 &=& e^{-i\int_0^t \theta_1^{(k)} d\tau H^{(k)} }
e^{-i  \int_0^t [z_{\mu }\theta_1^{(k)} + \frac{{\theta_0}^{(k)}}{\mu }] d\tau }  |\phi^{(k)} (0)\rangle\\
 &=& U^{(k)}(t)
e^{u^{(k)}(t) +i v^{(k)}(t)}  |\phi^{(k)} (0)\rangle\label{dn}
\end{eqnarray}
It is extremely important to realize that $z_\mu$ and $\mu$ are the {\it same\/} for all the subspaces indexed by $k$. This degeneracy of the eigenvalue problem was one of the crucial tricks that led in \citep{LC98} to the discovery of self-scattering solutions. The same is here, but the necessity of maintaining the degeneracy is one of the difficulties one encounters when trying to employ the method in arbitrary dimensions, and for arbitrary $H$.

Since $\mu$ and $z_\mu$ are independent of $k$, any linear combination 
$|\varphi(t)\rangle = \sum_k {\vartheta}^{(k)} |\varphi^{(k)}(t)\rangle$ fulfils
\begin{eqnarray}
z_{\mu }|\varphi(t) \rangle &=& \left(\frac{1}{\mu }\rho -  H\right )|\varphi(t) \rangle,\cr
i|\dot\varphi(t) \rangle &=& \frac{1}{\mu }f(\rho )|\varphi(t) \rangle ,\nonumber
\end{eqnarray}
and therefore can be used to build the projector $P$ from the dressing transformation,
\begin{eqnarray}
P(t) &=& \frac{ |\varphi(t)\rangle  \langle \varphi(t)
|}{\langle \varphi(t)|\varphi(t)\rangle }  \cr &=&  \frac{ \sum_{l,k}  {\vartheta}^{(k)} \overline{{\vartheta}^{(l)}}|\varphi^{(k)}(t)\rangle \langle \varphi^{(l)}(t)| }{\sum_k |{\vartheta}^{(k)}|^2\langle \varphi^{(k)}(t)|\varphi^{(k)}(t)\rangle}\cr &=& U(t) \sum_{l,k} c_{kl}(t) |\varphi^{(k)}(0)\rangle \langle \varphi^{(l)}(0)|U(t)^\dag,\nonumber
\end{eqnarray}
where $U(t)= \sum_k U^{(k)}(t)$,
\be
c_{kl}(t) =  \frac{  {\vartheta}^{(k)} \overline{{\vartheta}^{(l)}} e^{[u^{(k)}(t) + u^{(l)}(t)] + i [v^{(k)}(t) - v^{(l)}(t)]} }{\sum_i |{\vartheta}^{(i)}|^2  e^{2u^{(i)}(t)}}\label{c_kl},
\ee
and we sum over all the blocks. $U$ simultaneously generates the evolution of the seed solution $\rho(t)= U(t)\rho(0)U(t)^\dag$. The difference between $\rho(t)=\rho[0](t)$ and $\rho[1](t)$ is precisely in the presence of nontrivial $c_{kl}(t)$. The self-scattering phenomenon comes from $c_{kl}(t)$. The essence of
 finding a nontrivial dressing transformation is in guaranteeing that $c_{kl}(t)$ depend on time. Otherwise, the dynamics would be as linear as the one of the seed solution.

To construct the dressing transformation it is enough to find two-dimensional operators which fulfill (\ref{st}) for fixed $H$, $\mu$ and $z_{\mu}$.

The matrix representation of $\rho$ in the basis of eigenvectors of $H$ looks as follows:
\begin {equation}
\rho = \left(\begin{array}{cccc} \rho^{(1)}  & 0  & 0 &  \cdots\cr
0  & \rho^{(2)}  & 0 &  \cdots\cr
 0 & 0  & \rho^{(3)} &  \cdots\cr
\vdots  & \vdots  & \vdots &  \ddots \cr
\end{array}\right)\nonumber,
\end {equation}
\begin{eqnarray}
\rho^{(k)} = {\bf R}^{(k)} \rho
{\bf R}^{(k)} = \left(\begin{array}{cc} \rho_{1}^{(k)}  & |c^{(k)}|e^{i\gamma^{(k)}} \cr
|c^{(k)}|e^{-i\gamma^{(k)}}  &   \rho_{2}^{(k)} \cr
\end{array}\right)\nonumber
\end{eqnarray}
Eigenvalue problem (\ref{st}) implies two conditions:
\begin{eqnarray}
|c^{(k)}|^2 &=& (\rho_{1}^{(k)} - x -\alpha h_{1}^{(k)})(\rho_{2}^{(k)} - x - \alpha h_{2}^{(k)})
- (\beta h_{1}^{(k)} + y)(\beta h_{2}^{(k)} + y),\nonumber\\
0 &=&(\rho_{1}^{(k)} - x - \alpha h_{1}^{(k)})(\beta h_{2}^{(k)} + y) + (\rho_{2}^{(k)} - x - \alpha h_{2}^{(k)})(\beta h_{1}^{(k)} + y),\nonumber
\end{eqnarray}
where $\mu = \alpha + i\beta$ and $\mu z_{\mu} = x+iy$. Note that $\alpha$, $\beta$, $x$, $y$ are independent of $k$.
The above two conditions fix two parameters of the $2\times 2$ matrix $\rho^{(k)}$ as a function of $H^{(k)}$, $\mu$, $z_{\mu}$ and $\rho_{2}^{(k)}$:
\begin{eqnarray}
|c^{(k)}|^2 &=& -\left[\left(\frac{\rho_{2}^{(k)} - x - \alpha h_{2}^{(k)}}{\beta h_{2}^{(k)} + y}\right)^2 +1\right](\beta h_{1}^{(k)} + y)(\beta h_{2}^{(k)} + y),\label{pos}\\
\rho_{1}^{(k)} &=&  x + \alpha h_{1}^{(k)} - (\rho_{2}^{(k)} - x - \alpha h_{2}^{(k)})\frac{\beta h_{1}^{(k)} + y}{\beta h_{2}^{(k)} + y}.
\end{eqnarray}
The fourth parameter $\gamma^{(k)}$ remains undetermined. (\ref{pos}) implies that $(\beta h_{1}^{(k)} + y)(\beta h_{2}^{(k)} + y)$ has to be negative. It means that $-\frac{y}{\beta}$ is between $h_{1}^{(k)}$ and $h_{2}^{(k)}$, for any $k$.  It is easy to check that for any $H$ we can also set all the parameters in such a  way that $\rho$ will be positive and normalized.

For such a ${\rho^{(k)} }(0)$ the eigenvector from (\ref{st}) is given by
\be
 |\phi^{(k)} (0)\rangle = \frac{1}{\sqrt{\beta|h_1^{(k)}-h_2^{(k)}|}}
\left(\begin{array}{c} \sqrt{|\beta  h_2^{(k)} +y|}e^{i\gamma^{(k)}}    \cr
\sqrt{|\beta h_1^{(k)} +y|} \frac{i-w^{(k)}}{\sqrt{(w^{(k)})^2 +1}} \cr
\end{array}\right) \equiv \left(\begin{array}{c}  \phi_1^{(k)} (0)   \cr
\phi_2^{(k)} (0) \cr
\end{array}\right),\label{nad w^k}
\ee
where 
\be
w^{(k)} = -\frac{\rho_2^{(k)}-\alpha h_2^{(k)}-x}{\beta h_2^{(k)} + y}.\label{w^k}
\ee
Now, we can use an equivalent form of the dressing transformation  \citep{LC98,KCL}, $\rho[1] = \rho + 2i\beta [P,H]$, and obtain
\be
\rho[1](t) = U(t)\Big(\rho(0) + 2i\beta\sum_{l,k=1}^n  c_{kl}(t)\big[ |\varphi^{(k)}(0)\rangle \langle \varphi^{(l)}(0)|,H\big]\Big)U(t)^\dag.
\label{rho[1] kom}
\ee
The latter formula shows another condition one has to guarantee in order to make $\rho[1](t)$ qualitatively different from $\rho(t)$: the commutator at the right-hand side of (\ref{rho[1] kom}) must be nonzero.

Representing $\rho[1]$ in the bases of eigenvectors of $H$,
\begin {equation}
\rho[1] = \left(\begin{array}{cccc} \rho[1]^{(11)}  & \rho[1]^{(12)}  & \rho[1]^{(13)} &  \cdots\cr
\rho[1]^{(21)}  & \rho[1]^{(22)}  & \rho[1]^{(23)} &  \cdots\cr
\rho[1]^{(31)}  & \rho[1]^{(32)}  & \rho[1]^{(33)} &  \cdots\cr
\vdots  & \vdots  & \vdots &  \ddots 
\end{array}\right),
\end {equation}
we get the $2\times 2$ blocks of the form
\begin{eqnarray}
\rho[1]^{(kl)} &=&
2i\beta c_{kl}\left(\begin{array}{cc} \phi_1^{(k)} (0)\overline{\phi_1^{(l)} (0)}(h_1^{(l)}- h_1^{(k)})e^{\omega_{11}^{(kl)}}  & \phi_1^{(k)} (0)\overline{\phi_2^{(l)} (0)}(h_2^{(l)}- h_1^{(k)})e^{\omega_{12}^{(kl)}}\\
\phi_2^{(k)} (0)\overline{\phi_1^{(l)} (0)}(h_1^{(l)}- h_2^{(k)})e^{\omega_{21}^{(kl)}} & \phi_2^{(k)} (0)\overline{\phi_2^{(l)} (0)}(h_2^{(l)}- h_2^{(k)}) e^{\omega_{22}^{(kl)}}
\end{array}\right)\nonumber\\
&=&\frac{ 2i {\rm sgn}(\beta)c_{kl}}{\sqrt{( h_1^{(k)} -  h_2^{(k)})( h_1^{(l)} -  h_2^{(l)})}}\left(\begin{array}{cc} r^{(kl)}_{11}e^{i(\gamma^{(k)} - \gamma^{(l)})}  &
r^{(kl)}_{12} \frac{-(w^{(l)}+i)e^{i\gamma^{(k)} }}{\sqrt{(w^{(l)})^2 +1}}  \\
r^{(kl)}_{21}\frac{(i - w^{(k)})e^{-i\gamma^{(l)} }}{\sqrt{(w^{(k)})^2 +1}} & r^{(kl)}_{22} \frac{-(i - w^{(k)})(w^{(l)}+i)}{\sqrt{(w^{(l)})^2 +1}\sqrt{(w^{(k)})^2 +1}} 
\end{array}\right),\nonumber\\
\label{64}
\end{eqnarray}
with $c_{kl}$ given by (\ref{c_kl}), 
\be
r^{(kl)}_{ij} &=&  \sqrt{|\beta h_{\sigma(i)}^{(k)} + y||\beta h_{\sigma(j)}^{(l)} +y|}( h_j^{(l)} -  h_i^{(k)})e^{-i\omega_{ij}^{(kl)}},\\ 
\omega_{ij}^{(kl)} &=& \int_0^t(\theta_1^{(k)}h^{(k)}_i- \theta_1^{(l)}h^{(l)}_j) d\tau,
\ee 
$\sigma(i)$ an odd permutation of $i\in\{ 1,2\}$, and
\begin{eqnarray}
\rho[1]^{(kk)} &=&
\left(\begin{array}{cc} \alpha h_1^{(k)} +x +w^{(k)}s^{(k)}  & s^{(k)}(1+\frac{2i}{w^{(k)}-i}c_{kk})e^{i\gamma^{(k)}}e^{-i\omega_{12}^{(kk)}}  \\
s^{(k)}(1-\frac{2i}{w^{(k)}+i}c_{kk})e^{-i\gamma^{(k)}}e^{-i\omega_{21}^{(kk)}} & \rho_2^{(k)}  
\end{array}\right),\nonumber\\\label{67}
\end{eqnarray}
where $s^{(k)}=\sqrt{((w^{(k)})^2 +1)|\beta  h_1^{(k)} +y||\beta  h_2^{(k)} +y|}$.

All these blocks can consist of nonzero elements, leading to the coupling between all the pairs of species in a given population. Note that the construction does not depend on the number of species. 

\subsection{Example: $3\times 3$ $\rho[1](t)$ constructed from a single $2\times 2$ seed block}

We are in position to build a wide variety of solutions to a large class of problems. Of course, the formulas describing a general case are quite complicated.
So, let us simplify the discussion by setting $\mu = i\beta$, $\mu z_{\mu} = iy$ (thus $\alpha=x=0$). Now $\beta h_{1}^{(k)} + y>0$, $\beta h_{2}^{(k)} + y<0$ and we obtain a simple $2\times 2$ block
\begin{eqnarray}
\rho^{(k)} = \left(
 \begin{matrix}
 0&\sqrt{-(\beta h_{1}^{(k)} + y)(\beta h_{2}^{(k)} + y)}e^{i\gamma^{(k)}}\\
 \sqrt{-(\beta h_{1}^{(k)} + y)(\beta h_{2}^{(k)} + y)}e^{-i\gamma^{(k)}} &0
 \end{matrix}
 \right).\nonumber
\end{eqnarray}
which is Hermitian but not positive.
We can use this  block for constructing the transformation in an even-dimensional case. Let  us skip the index $k$. We can make the seed solution odd-dimensional by adding the 1-dimensional block
$\rho^{(2)} = 0$, implying  $y=-\beta h_3 $.  We assume $\beta>0$, therefore $h_{1}>h_{3}>h_{2}$. Under all  these assumptions our seed solution becomes
\begin{eqnarray}
\rho = \left(
 \begin{matrix}
 0&\sqrt{\beta^2 (h_1 - h_3)(h_3 - h_2)}e^{i\phi_1}&0\\
 \sqrt{\beta^2 (h_1 - h_3)(h_3 - h_2)}e^{-i\phi_1} &0&0\\
 0&0&0
 \end{matrix}
 \right).\nonumber
\end{eqnarray}
To make  our example yet more explicit we use  $f(\rho)=  F(t)\rho^2$ for which our $\rho$ is a constant solution because of $[H,\rho^2 ]=0$. The dressing  transformation produces
\begin{eqnarray}
\rho[1]
  &=& \beta \sqrt{ (h_1 - h_3)(h_3 - h_2)}\times \nonumber \cr
 &\times& \left(
 \begin{matrix}
  0& -e^{i\phi_1}(1-\frac{2 |a_2|^2 }{\chi(t)} )&\frac{-2ie^{\psi(t) }  a_1\overline{a_2}e^{i\phi_1}\sqrt{h_1 - h_3}}{\chi(t)\sqrt{ h_1 - h_2}}\\
  -e^{-i\phi_1}(1-\frac{2 |a_2|^2 }{\chi(t)} )&0&\frac{-2e^{\psi(t) }  a_1\overline{a_2}\sqrt{h_3 - h_2}}{\chi(t)\sqrt{ h_1 - h_2}}\\
\frac{2ie^{\psi(t) }  a_2\overline{a_1}e^{-i\phi_1}\sqrt{h_1 - h_3}}{\chi(t)\sqrt{ h_1 - h_2}}&\frac{-2e^{\psi(t) }  a_2\overline{a_1}\sqrt{h_3 - h_2}}{\chi(t)\sqrt{ h_1 - h_2}}&0
 \end{matrix}
 \right), \nonumber
\end{eqnarray}
where $\psi(t) = -\beta (h_1 - h_3)( h_3 - h_2)\int_0^t F(s)ds$ and $\chi(t) = |a_1|^2e^{ 2\psi(t) }+|a_2|^2$.

Now, we use this simple traceless and non-positive $3\times 3$ matrix to construct a density matrix (i.e. a positive solution with $\Tr\rho=1$). Let us first see what happens if we just add to $\rho$ a normalized multiple of $I$:  $\sigma = \rho +\frac{1}{3}I$, $\Tr\rho=0$, $\Tr\sigma=1$. The equation is
\be
i\dot\rho &=& i\dot\sigma =  [H,g(\sigma)] = [H,\theta_2 \sigma^2 + \theta_1 \sigma + \theta_0 I] = [H,\theta_2 \sigma^2 + \theta_1 \sigma ]\cr
 &=& [H,\theta_2 (\rho +\frac{1}{3}I)^2 + \theta_1 (\rho +\frac{1}{3}I) ] = [H,\theta_2 \rho^2 + (\theta_1 +\frac{2}{3}\theta_2)\rho ]  =  [H,f(\rho)]\label{r} \nonumber
\ee
where we used again the theorem from Appendix,  describing $f$ and $g$  as a polynomial of a matrix. So, if we want to find a solution of  $ i\dot{\sigma} =  [H,g(\sigma)]$ for a density matrix we should find a traceless solution of $i\dot{\rho}= [H,\theta_2 \rho^2 + (\theta_1 +\frac{2}{3}\theta_2)\rho ]$.

 We know how to find a traceless solution of the problem with $f(\varrho) = F(t)\varrho^2$, but it is easy to verify that $\rho= \exp(-i\int_0^t(\theta_1 +\frac{2}{3}\theta_2)d\tau H )\varrho \exp(i\int_0^t(\theta_1 +\frac{2}{3}\theta_2)d\tau H )$ is a solution of $i\dot{\rho}= [H,\theta_2 \rho^2 + (\theta_1 +\frac{2}{3}\theta_2)\rho ]$  if $\varrho$ is a solution of $i\dot{\varrho} = [H,\theta_2 \varrho^2]$.

For our concrete example $g$ and $H$ are fixed, and we know that the eigenvalues of $\sigma$ are:
$\lambda_{1} = \frac{1}{3} + \sqrt{\beta^2 (h_1 - h_3)( h_3 - h_2)}\equiv \frac{1}{3} + T$, $\lambda_{2} = \frac{1}{3} - \sqrt{\beta^2 (h_1 - h_3)( h_3 - h_2)} \equiv \frac{1}{3} - T$, $\lambda_{3} = \frac{1}{3}$. Because we are interested in positive matrices we select parameters so that $T<\frac{1}{3}$.
The coefficients $\theta_1$ and $\theta_2$ can be computed from
\be
\theta_1 &=& \frac{(\lambda_2+\lambda_1)(g(\lambda_2)- g(\lambda_1))}{(\lambda_2-\lambda_1)(\lambda_2-\lambda_3)} - \frac{(\lambda_3+\lambda_1)(g(\lambda_3)- g(\lambda_1))}{(\lambda_3-\lambda_1)(\lambda_2-\lambda_3)}\nonumber\cr
&=& \frac{g(\lambda_2)- g(\lambda_1)}{3T^2} - \frac{(\frac{2}{3} + T)(g(\lambda_3)- g(\lambda_1))}{T^2}\nonumber,\cr
\theta_2 &=& \frac{g(\lambda_2)- g(\lambda_1)}{(\lambda_2-\lambda_1)(\lambda_2-\lambda_3)} - \frac{g(\lambda_3)- g(\lambda_1)}{(\lambda_3-\lambda_1)(\lambda_2-\lambda_3)}= \frac{g(\lambda_2)+ g(\lambda_1)-2g(\lambda_3)}{2T^2}\nonumber,
\ee
and then
\be
\theta_1 +\frac{2}{3}\theta_2 &=& \big(g(\lambda_2)- g(\lambda_1)\big) \frac{2}{3T^2} + \big(g(\lambda_1)- g(\lambda_3)\big)\frac{\frac{4}{3} + T}{T^2}.
\ee
As a result, we obtain the density matrix which solves $i\dot\rho =  [H,g(\rho)]$ for arbitrary $H$ and $g$:
\be
\rho =
\left(
 \begin{matrix}
 \frac{1}{3}&A&B\\
 \bar{A} & \frac{1}{3}&C\\
 \bar{B}&\bar{C}& \frac{1}{3}
 \end{matrix}
 \right)\nonumber
 \ee
 where 
 \be
 A &=& -e^{i\phi_1}\beta \sqrt{ (h_1 - h_3)(h_3 - h_2)}\left(1-\frac{2 |a_2|^2 }{\chi(t)} \right)e^{i\kappa(h_1 - h_2)},\\
 B &=&  \frac{-2ie^{\psi(t) }e^{i\kappa(h_1 - h_3)}\beta  a_1\overline{a_2}e^{i\phi_1}(h_1 - h_3)\sqrt{h_3 - h_2}}{\chi(t)\sqrt{ h_1 - h_2}},\\
 C&=& \frac{-2e^{\psi(t) } e^{i\kappa(h_2 - h_3)}\beta a_1\overline{a_2}(h_3 - h_2)\sqrt{h_1 - h_3}}{\chi(t)\sqrt{ h_1 - h_2}},
 \ee 
with $\kappa = -\int_0^t(\theta_1 +\frac{2}{3}\theta_2)d\tau$, $\psi(t) = -\beta (h_1 - h_3)( h_3 - h_2)\int_0^t \theta_2(s)ds$. The general form of solution we have given in the previous section is much more complex and flexible from the point of view of modeling, but our simplified example is easier for further analysis. 

Let us finally write the associated set of rate equations: 
\be
\dot{p}_{12}  &=& \Big[(\theta_1 +\frac{2}{3}\theta_2)(p'_{12}-\frac{1}{3}) +\theta_2\Big((p'_{13}- \frac{1}{3})(p_{23}- \frac{1}{3}) - (p_{13}- \frac{1}{3})(p'_{23}-\frac{1}{3})\Big)\Big](h_1 - h_2)\nonumber\\
\dot{p'}_{12} &=& \Big[\theta_2\Big((p_{13}-\frac{1}{3})(p_{23}- p) + (p'_{13}-\frac{1}{3})(p'_{33}- p)\Big) - (\theta_1 +\frac{2}{3}\theta_2)(p_{12}-\frac{1}{3})\Big](h_1 - h_2)\nonumber\\
\dot{p}_{13} &=& \Big[(\theta_1 +\frac{2}{3}\theta_2)(p'_{13}-\frac{1}{3})+\theta_2\Big((p_{12}-\frac{1}{3})(p'_{23}-\frac{1}{3}) - (p'_{12}-\frac{1}{3})(p_{23}-\frac{1}{3})\Big)\Big](h_1 - h_3)\nonumber\\
\dot{p'}_{13} &=& \Big[\theta_2\Big((p_{12}-\frac{1}{3})(p_{23}-\frac{1}{3}) - (p'_{12}-\frac{1}{3})(p'_{23}-\frac{1}{3})\Big)- (\theta_1 +\frac{2}{3}\theta_2)(p_{13}-\frac{1}{3}) \Big](h_1 - h_3)  \nonumber\\
\dot{p}_{23} &=& \Big[(\theta_1 +\frac{2}{3}\theta_2)(p'_{23}-\frac{1}{3})+\theta_2\Big((p_{12}-\frac{1}{3})(p'_{13}-\frac{1}{3}) - (p'_{12}-\frac{1}{3})(p_{13}-\frac{1}{3})\Big)\Big](h_2 - h_3)\nonumber\\
\dot{p'}_{23} &=& \Big[\theta_2\Big((p_{12}-\frac{1}{3})(p_{13}-\frac{1}{3}) - (p'_{12}-\frac{1}{3})(p'_{13}-\frac{1}{3})\Big)- (\theta_1 +\frac{2}{3}\theta_2)(p_{23}-\frac{1}{3}) \Big](h_2 - h_3)  \nonumber
\ee
Since $C = i\sqrt{\frac{h_3-h_2}{h_1-h_3}}Be^{i[\kappa(h_2 - h_1) - \phi_1]}$ one can replace this system of six equations for catalytic processes by four equations for $p_{12}$, $p_{13}$, $p'_{12}$, $p'_{13}$,  involving auto-catalytic terms and new time dependent coefficients. The form of the rate equations explains why the eigenvalues of $H$ effectively encode the kinetic constants of the dynamics, here $k_{ij}=h_i-h_j$. 

\subsubsection{$g(\rho) = \rho \sin\rho$}

\begin{figure}
\centering
\includegraphics[width=.5\textwidth]{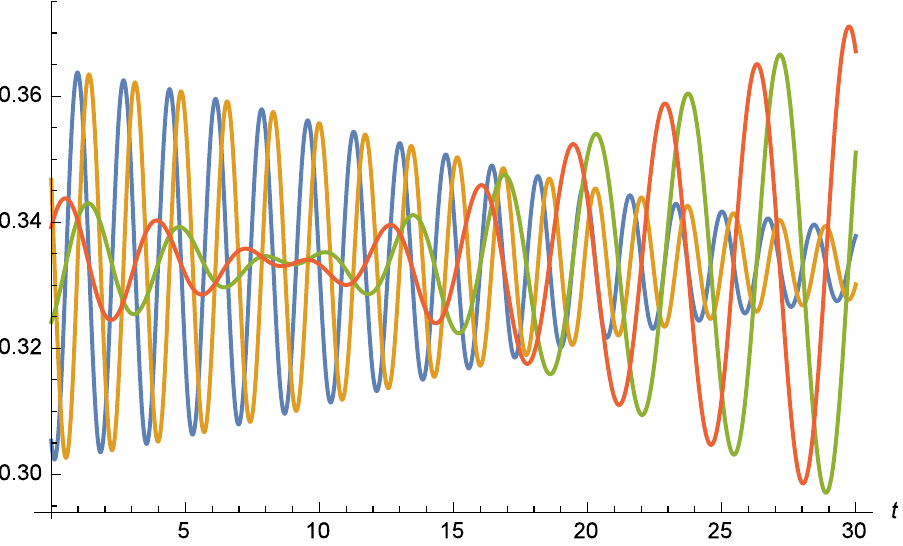}
\includegraphics[width=.5\textwidth]{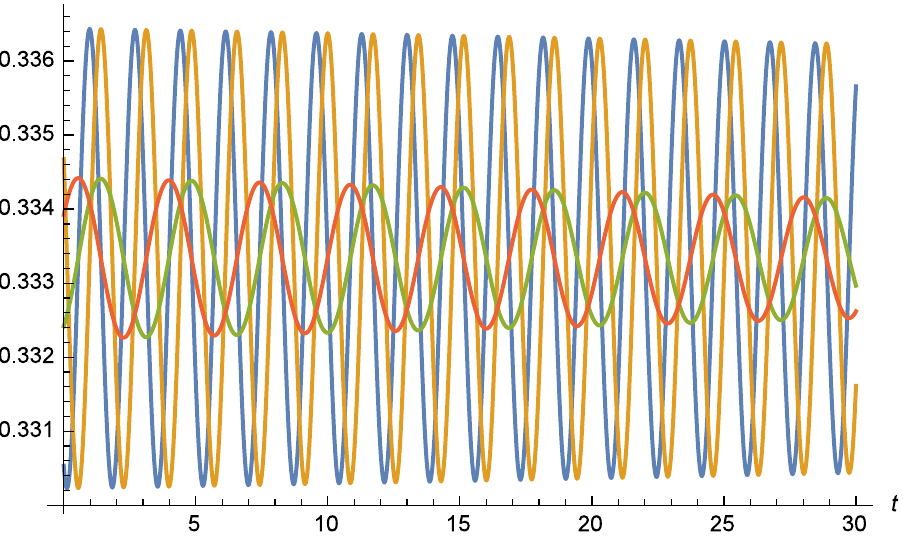}
\includegraphics[width=.5\textwidth]{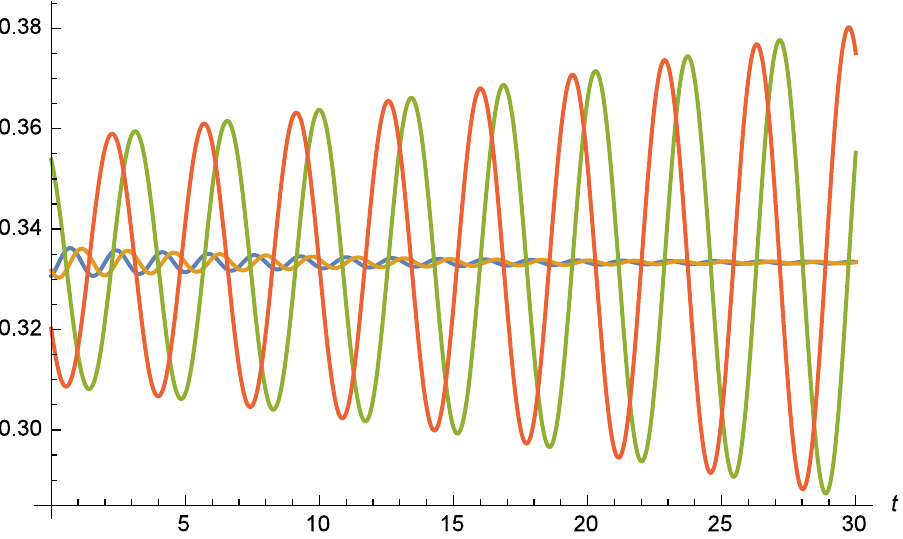}
\caption{Dynamics of the four populations $p_{12}$ (green), $p_{13}$ (blue), $p'_{12}$ (red), $p'_{13}$ (yellow), for time-independent $g_1(\rho) = \rho \sin(\rho)$. The top plot: $h_1=2$, $h_2=1$, $h_3=4$,  $a_1=1+2i$, $a_2=2+3i$, $\beta=0.01$, $\phi_1=1$. In the lower plots we change a single parameter with respect to the top one: $\beta=0.001$ (middle), $a_2=0.1$ (lowest). The pairs red-green and blue-yellow exhibit a typical predator-prey, or species-resources Volterra-type shift of oscillation. The system can be interpreted as consisting of two species functioning in two niches, which are nevertheless not completely isolated from one another. The reduction from six to four variables, for the price of making some coefficients time dependent, turns the remaining two populations $p_{23}$, $p'_{23}$ into an effective environment for $p_{12}$, $p_{13}$, $p'_{12}$, $p'_{13}$.}
\label{Fig2}
\end{figure}

Take  $g(\rho) =g_1(\rho) = \rho \sin\rho$, 
\be
i\dot\rho &=&[H,\rho \sin\rho].
\ee
The function itself is time independent, so after explicit integrations one finds
\be
\psi_1(t) &=&  -\frac{t}{2\beta}
\Bigg(\left(\frac{1}{3} + T\right)\sin\left(\frac{1}{3} + T\right) + \left(\frac{1}{3} - T\right)\sin\left(\frac{1}{3} - T\right) -\frac{2}{3} \sin\frac{1}{3}\Bigg)\nonumber\\
\kappa_1(t)  &=&-\frac{2t}{3T^2}\Bigg(\left(\frac{1}{3} - T\right)\sin\left(\frac{1}{3} - T\right) - \left(\frac{1}{3} + T\right)\sin\left(\frac{1}{3} + T\right)\Bigg)\nonumber\\
&\pp=&-\frac{(4/3+T)t}{T^2}\Bigg(\left(\frac{1}{3} + T\right)\sin\left(\frac{1}{3} + T\right) -\frac{1}{3}\sin\frac{1}{3} \Bigg)\nonumber.
\ee
$p_{12}$, $p_{13}$, $p'_{12}$, $p'_{13}$ are plotted in Fig.~\ref{Fig2}.

\subsubsection{$g(\rho) = \dot\omega(t)\rho \sin(\omega(t)\rho)$}

\begin{figure}
\centering
\includegraphics[width=.5\textwidth]{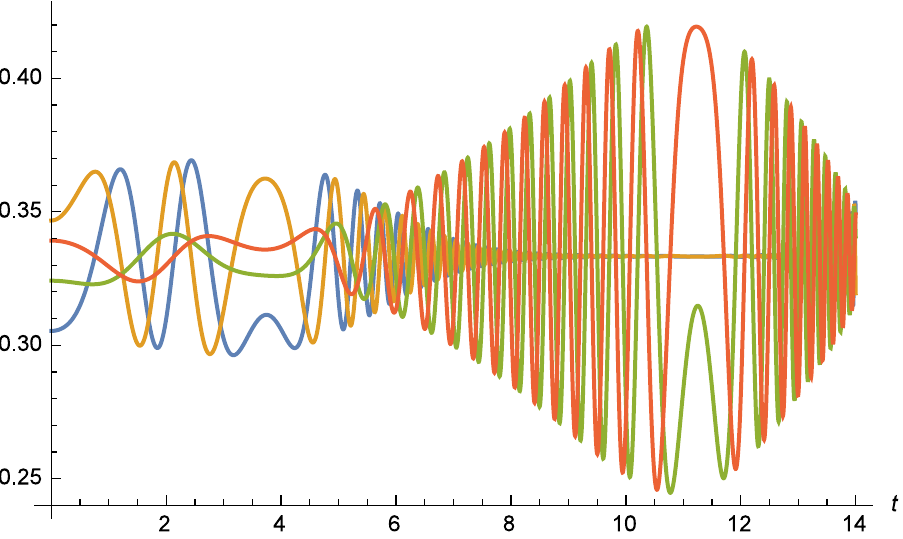}
\includegraphics[width=.5\textwidth]{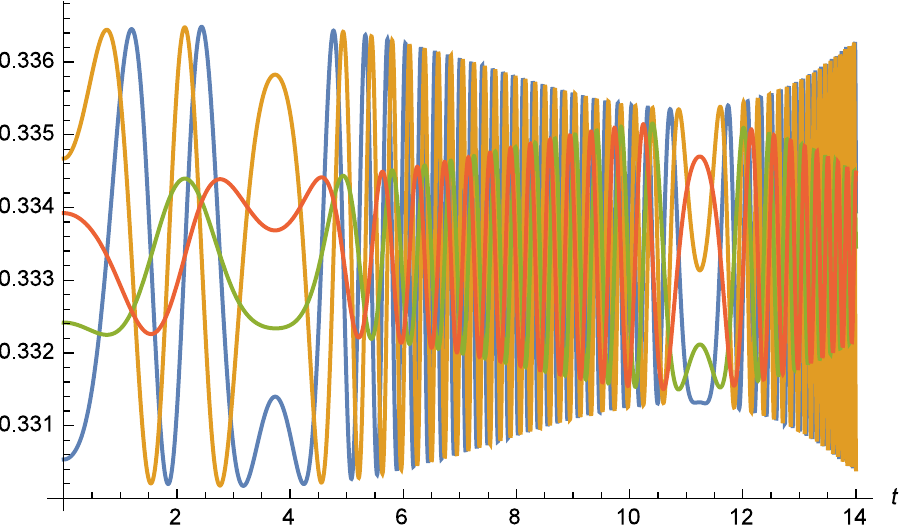}
\includegraphics[width=.5\textwidth]{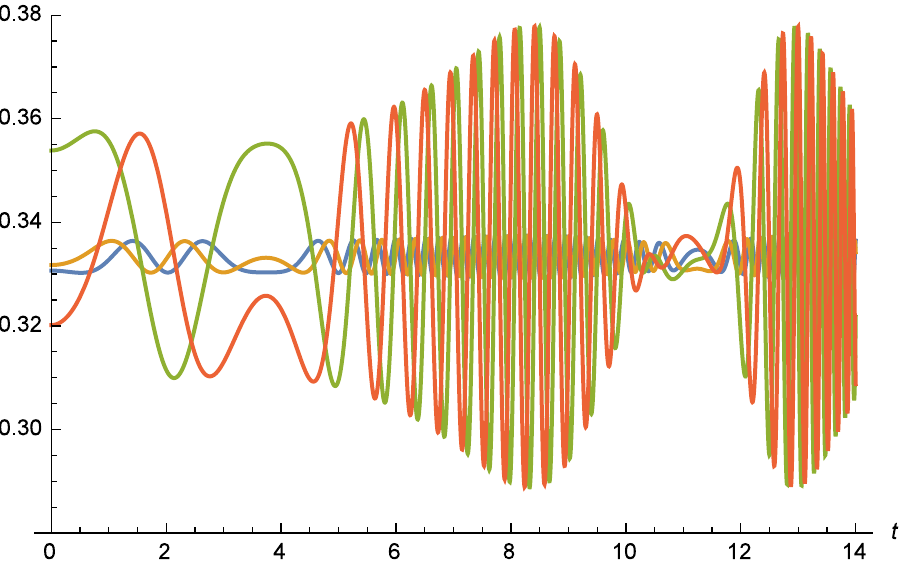}
\caption{$g_2(\rho) = \dot{\omega(t)}\rho \sin(\omega(t)\rho)$ for $\omega(t)=\omega t$, $\omega=1$. The remaining parameters the same as in Fig.~\ref{Fig2}.}
\label{Fig3}
\end{figure}

The second example involves an explicitly time dependent function  $g(\rho) = g_2(\rho) = \dot\omega(t)\rho \sin(\omega(t)\rho)$. 
Then 
\be
i\dot\rho &=&[H,\dot\omega(t)\rho \sin(\omega(t)\rho)].
\ee
After integration,
\be
\psi_2(t) &=&  \frac{1}{2\beta}\Bigg(\cos\left[\left(\frac{1}{3} + T\right) \omega(t)\right] + \cos\left[\left(\frac{1}{3} - T\right) \omega(t)\right] -2\cos\frac{\omega(t)}{3}  \Bigg)\nonumber\\
\kappa_2(t)  &=&\frac{2}{3T^2}\Bigg(\cos\left[\left(\frac{1}{3} - T\right)\omega(t)\right] - \cos\left[\left(\frac{1}{3} + T\right)\omega(t)\right]\nonumber\\
&+&\frac{4/3+T}{T^2}\Bigg(\cos\left[\left(\frac{1}{3} + T\right)\omega(t)\right] -\cos\frac{\omega(t)}{3}\Bigg)\nonumber.
\ee
Let us see what kind of a dynamics one gets for the simplest, linear time dependence $\omega(t)=\omega t$. The  four kinetic variables are shown at Fig.~\ref{Fig3}.

\subsection{Dynamics of 30 interacting species:  Seed $\rho$ with three $2\times 2$ blocks}

In order to obtain a more complicated solution, but still of a reasonably compact form, we put $w^{(k)}=0$ in the general formula (\ref{w^k}). The two-dimensional blocks are as follows
\begin{eqnarray}
\rho^{(k)} = \left(
 \begin{matrix}
 x+\alpha h^{(k)}_{1}&\sqrt{|\beta h^{(k)}_{1} + y||\beta h^{(k)}_{2} + y|}e^{i\phi^{(k)}}\\
 \sqrt{|\beta h^{(k)}_{1} + y||\beta h^{(k)}_{2} + y|}e^{-i\phi^{(k)}} &x+\alpha h^{(k)}_{2}.
 \end{matrix}
 \right),\nonumber
\end{eqnarray}
with the eigenvalues
\be
\lambda^{(k)}_{1} &=& x + \frac{\alpha(h^{(k)}_{1}+ h^{(k)}_{2})+\sqrt{\alpha^2(h^{(k)}_{1}- h^{(k)}_{2})^2+4|\beta h^{(k)}_{1} + y||\beta h^{(k)}_{2} + y|}}{2},\nonumber\\
\lambda^{(k)}_{2} &=& x + \frac{\alpha(h^{(k)}_{1}+ h^{(k)}_{2})-\sqrt{\alpha^2(h^{(k)}_{1}- h^{(k)}_{2})^2+4|\beta h^{(k)}_{1} + y||\beta h^{(k)}_{2} + y|}}{2}\nonumber.
\ee
Recall that $\alpha$, $\beta$, $x$, $y$ are $k$-independent. 
Now, take positive $x$, $\alpha$, and $h_i^{(k)}$. Then ${\rm Tr}(\rho^{(k)})= 2x +\alpha(h^{(k)}_{1}+ h^{(k)}_{2})>0$. Because in this case $y\beta<0$,
\be
{\rm det}(\rho^{(k)})&=& (x+\alpha h^{(k)}_{1})(x+\alpha h^{(k)}_{2})-|\beta h^{(k)}_{1} + y||\beta h^{(k)}_{2} + y|\nonumber\\
&=& x^2 +y^2+h^{(k)}_{1} h^{(k)}_{2}(\alpha^2+\beta^2) +(x\alpha+y\beta)(h^{(k)}_{1}+ h^{(k)}_{2})\nonumber,
\ee
so $\rho^{(k)}$ becomes positive if $x\alpha>|y\beta|$.

The solution we produce from this seed solution is a little less complicated than the general one. For $k\neq l$,
\begin{eqnarray}
\rho[1]^{(kl)} &=&
\frac{ 2i {\rm sgn}(\beta)c_{kl}}{\sqrt{( h_1^{(k)} -  h_2^{(k)})( h_1^{(l)} -  h_2^{(l)})}}\left(\begin{array}{cc} r^{(kl)}_{11}e^{i(\gamma^{(k)} - \gamma^{(l)})}  &
-ir^{(kl)}_{12} e^{i\gamma^{(k)} } \nonumber \cr
ir^{(kl)}_{21}e^{-i\gamma^{(l)} } & r^{(kl)}_{22} \nonumber\cr
\end{array}\right)\nonumber,
\end{eqnarray}
where 
\be
r^{(kl)}_{ij} &=&  \sqrt{|\beta h_{\sigma(i)}^{(k)} + y||\beta h_{\sigma(j)}^{(l)} +y|}( h_j^{(l)} -  h_i^{(k)})e^{-i\omega_{ij}^{(kl)}},\nonumber\\
\omega_{ij}^{(kl)} &=& \int_0^t(\theta_1^{(k)}h_i^{(k)}- \theta_1^{(l)}h_j^{(l)}) d\tau,\nonumber
\ee
(again, $\sigma(i)$ is an odd permutation of $i\in\{ 1,2\}$). For $k=l$,
\begin{eqnarray}
\rho[1]^{(kk)} &=&
\left(\begin{array}{cc} \alpha h_1^{(k)} +x   & s^{(k)}(1-2c_{kk})e^{i\gamma^{(k)}}e^{-i\omega_{12}^{(kk)}}  \cr
s^{(k)}(1-2c_{kk})e^{-i\gamma^{(k)}}e^{-i\omega_{21}^{(kk)}} & \alpha h_1^{(k)} +x   \cr
\end{array}\right),\nonumber
\end{eqnarray}
where $s^{(k)}=\sqrt{|\beta  h_1^{(k)} +y||\beta  h_2^{(k)} +y|}$,
$c_{kl} =  \frac{  {\vartheta}^{(k)} \overline{{\vartheta}^{(l)}} e^{[u^{(k)}(t) + u^{(l)}(t)] + i [v^{(k)}(t) - v^{(l)}(t)]} }{\sum_i |{\vartheta}^{(i)}|^2  e^{2u^{(i)}(t)}}$. 
Explicit forms of $u^{(k)}(t)$ and $v^{(k)}(t)$ are given by
\begin{eqnarray}
u^{(k)}(t) +i v^{(k)}(t)
&=& \frac{\alpha[ f(\lambda^{(k)}_2)(\lambda^{(k)}_1 -x) + (x - \lambda^{(k)}_2) f(\lambda^{(k)}_1 )]    +\beta y  [f(\lambda^{(k)}_1 )
- f(\lambda^{(k)}_2 )]}{(\alpha^2+\beta^2)(\lambda^{(k)}_1 - \lambda^{(k)}_2) } \nonumber\\
&+&i \frac{y\alpha[  f(\lambda^{(k)}_1 )
- f(\lambda^{(k)}_2 )] -\beta[f(\lambda^{(k)}_2)(\lambda^{(k)}_1 -x) + (x - \lambda^{(k)}_2) f(\lambda^{(k)}_1 )]     }{(\alpha^2+\beta^2)(\lambda^{(k)}_1 - \lambda^{(k)}_2) } \nonumber
\end{eqnarray}
Now take a seed $\rho$ consisting of three two-dimensional blocks. 
$\rho[1]$ is $6\times 6$, and consists of nine, nonzero $2\times 2$ blocks $\rho[1]^{(kl)}$, $k,l=1,2,3$. A practical advice in combining the blocks is to keep in mind that the denominators in $c_{kl}$ are the same for all $k,l$, and involve the sum over $k$ from $k=1$ to $k=3$. 
For example,
$$
c_{12} =  \frac{  {\vartheta}^{(1)} \overline{{\vartheta}^{(2)}} e^{[u^{(1)}(t) + u^{(2)}(t)] + i [v^{(1)}(t) - v^{(2)}(t)]} }
{ |{\vartheta}^{(1)}|^2  e^{2u^{(1)}(t)}+|{\vartheta}^{(2)}|^2  e^{2u^{(2)}(t)}+|{\vartheta}^{(3)}|^2  e^{2u^{(3)}(t)}}.
$$
Note the presence of $|{\vartheta}^{(3)}|^2  e^{2u^{(3)}(t)}$ even though the block itself is indexed by 1,2. 
Secondly, the contribution from $c_{kl}$ enters all the matrix elements of $\rho[1]$, with the exception of the main diagonal. This is why all the species described by $\rho[1]$ interact with one another.

In order to obtain a density matrix we select $\alpha$ in such a way that
$\alpha\sum_{k=1}^3 (h^{(k)}_{1}+ h^{(k)}_{2} )<1$, then we put $x = \frac{1-\alpha\sum_{k=1}^3 (h^{(k)}_{1}+ h^{(k)}_{2})}{2}$.

\subsubsection{$f(\rho)= F(t)\rho^2$}

\begin{figure}
\centering
\includegraphics[width=.5\textwidth]{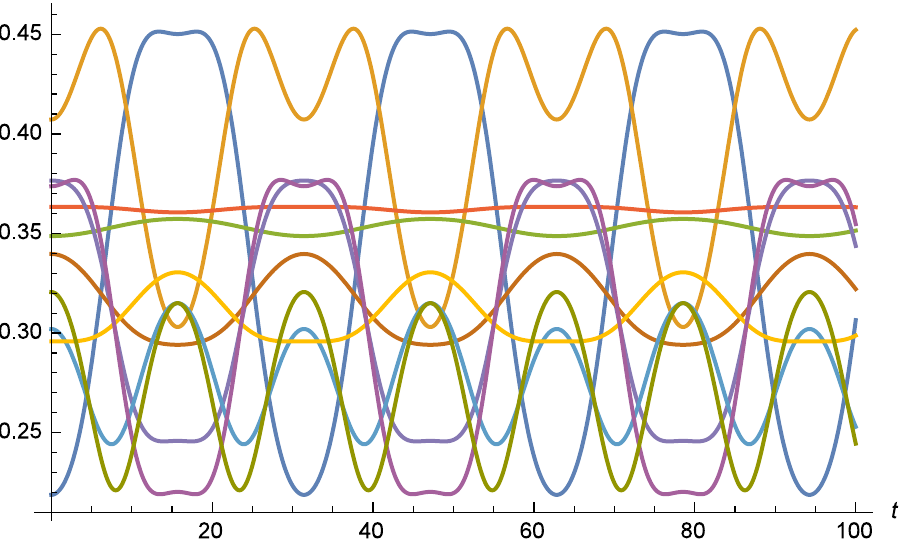}
\includegraphics[width=.5\textwidth]{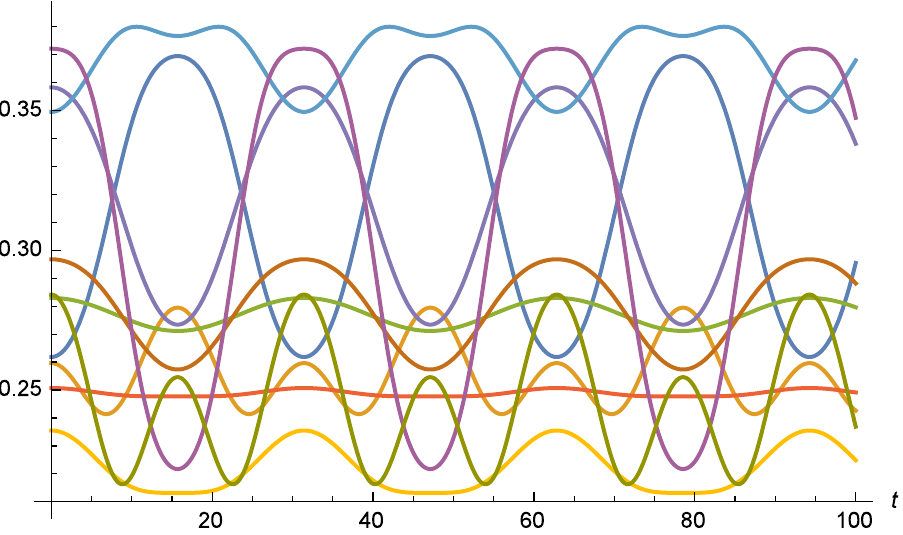}
\includegraphics[width=.5\textwidth]{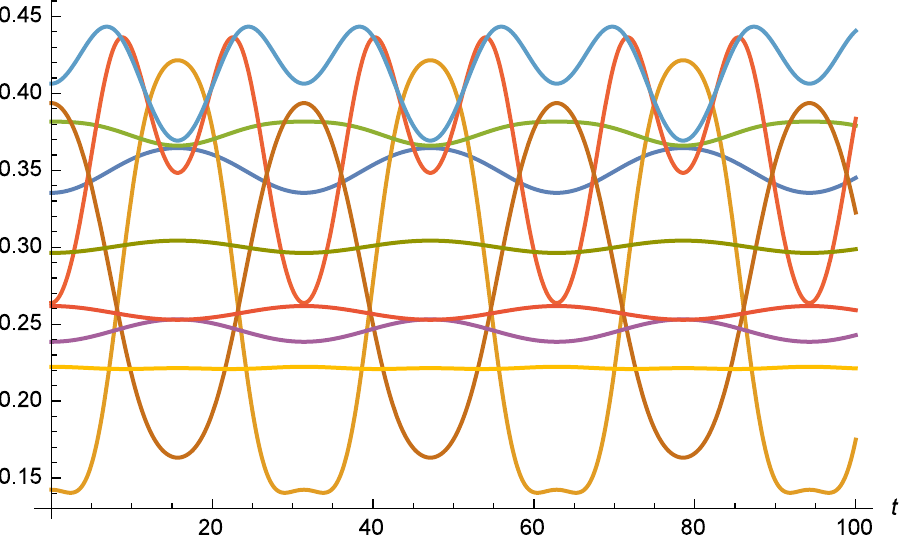}
\caption{30 populations for  $f(\rho) = \sin(\omega t) \rho^2$, $h_1=0.8$, $h_2=0.4$, $h_3=0.7$, $h_4=0.3$, $h_5=0.6$, $h_6=0.2$, $\vartheta^{(1)}=1+i$, $\vartheta^{(2)}=2+i$, $\vartheta^{(3)}=3+i$, $\alpha=0.2$, $\beta=1$, $\omega=0.2$, $\gamma_1=20$, $\gamma_2=30$, $\gamma_3=40$. The solution $\rho$ is normalized by $\Tr\rho=1$. Note that the sum of all the 30 probabilities exceeds 1. This shows that the probabilities correspond to several maximal sets, involving simultaneously different contexts for different collections of populations.}
\label{Fig6}
\end{figure}

\begin{figure}
\centering
\includegraphics[width=.5\textwidth]{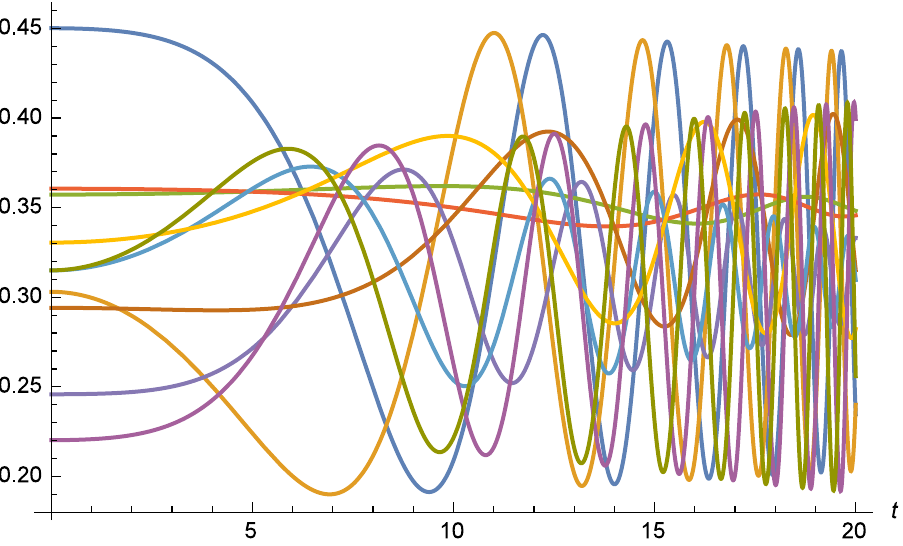}
\includegraphics[width=.5\textwidth]{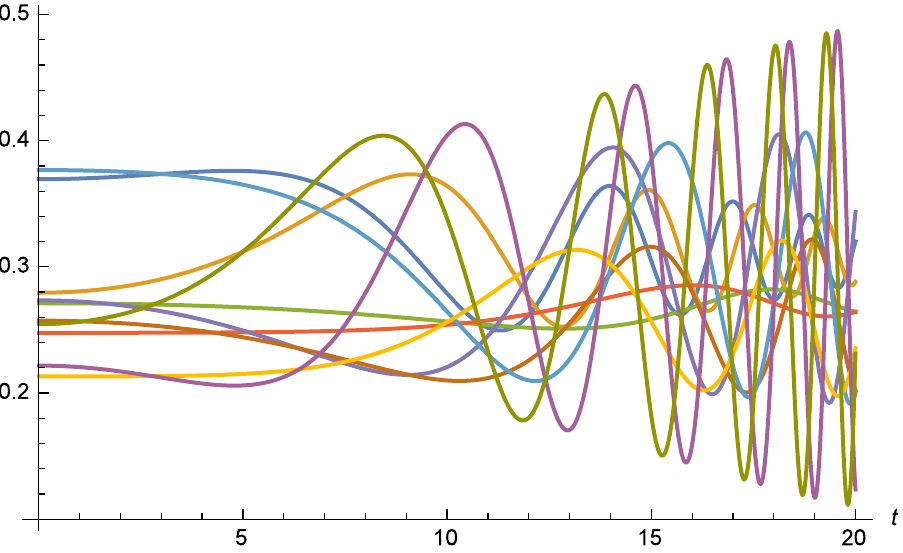}
\includegraphics[width=.5\textwidth]{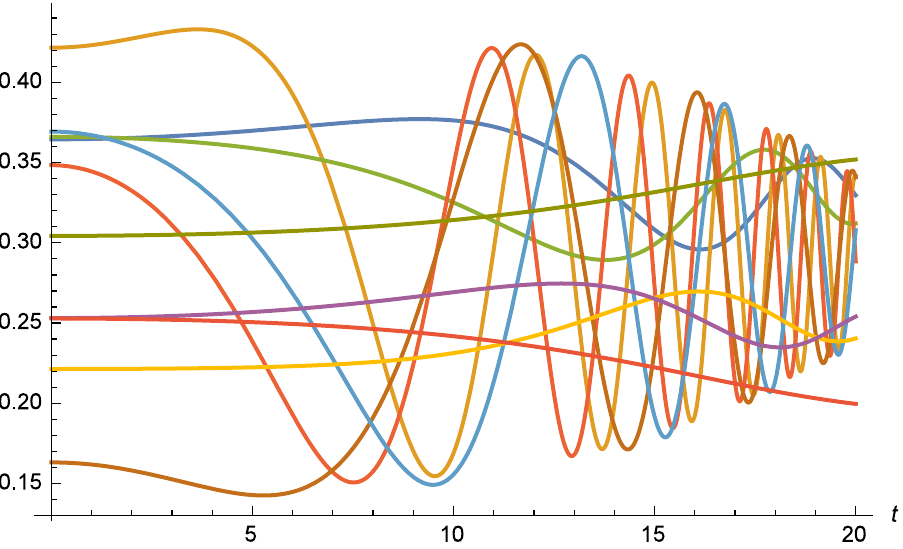}
\caption{30 populations for $f(\rho) = \sinh(\omega t) \rho^2$, the remaining parameters as in Fig.~\ref{Fig6}. Note the change of time scale. For short times the evolutions are similar for both $\sin \omega t$ and $\sinh \omega t$.}
\label{Fig11}
\end{figure}

Let us again assume that the coupling with environment is given by some explicit time-dependent function, say $f(\rho)= F(t)\rho^2$. Then
$ \theta_1^{(k)} = F(t)(\lambda^{(k)}_1+ \lambda^{(k)}_2)$,   $ \theta_0^{(k)}=-F(t)\lambda^{(k)}_1 \lambda^{(k)}_2$,
\be
u^{(k)}(t) +i v^{(k)}(t) &=&
-i  \int_0^t [z_{\mu }\theta_1^{(k)} + \frac{\theta_0^{(k)}}{\mu }] d\tau \nonumber \\
&=& \int_0^t F(\tau) d\tau  \left[\frac{\beta(y^2-x^2) +2\alpha yx }{\alpha^2+\beta^2} + \beta h^{(k)}_{1}h^{(k)}_{2} + y(h^{(k)}_{1} + h^{(k)}_{2} ) \right]\nonumber\\
&\pp=&+i\int_0^t F(\tau) d\tau  \left[\frac{\alpha (x^2-y^2) + 2\beta yx}{\alpha^2+\beta^2} -\alpha h^{(k)}_{1}h^{(k)}_{2}\right]\nonumber,
\ee
\be
c_{kl} &=&
 \frac{  {\vartheta}^{(k)} \overline{{\vartheta}^{(l)}} e^{\int_0^t F(\tau) d\tau[ \beta (h^{(k)}_{1}h^{(k)}_{2}+h^{(l)}_{1}h^{(l)}_{2}) + y(h^{(k)}_{1} + h^{(k)}_{2}+h^{(l)}_{1} + h^{(l)}_{2} )-i\alpha( h^{(k)}_{1}h^{(k)}_{2}- h^{(l)}_{1}h^{(l)}_{2}) ] }}{\sum_i |{\vartheta}^{(i)}|^2  e^{\int_0^t F(\tau) d\tau[ 2\beta h^{(i)}_{1}h^{(i)}_{2} + 2y(h^{(i)}_{1} + h^{(i)}_{2} )] }}\nonumber\\
\omega_{ij}^{(kl)} &=& \int_0^t F(\tau) d\tau\left[2x(h_i^{(k)}- h_j^{(l)})+\alpha[(h^{(k)}_{1} + h^{(k)}_{2} )h_i^{(k)}- (h^{(l)}_{1} + h^{(l)}_{2} )h_j^{(l)}]\right]\nonumber
\ee
The associated set of rate equations involves of 36 populations, but clarity of presentation will not be increased by writing down all the 30 time-dependent functions (the `diagonal' populations $p_{1}, \dots,p_{6}$ are constant if one works in the basis of eigenvectors of $H$). Yet, in order to get some flavor of the solution let us show at least some of them:
\be
p_{12}
&=& x+\frac{\alpha}{2}(h^{(1)}_{1}+ h^{(1)}_{2})+ \sqrt{|\beta  h_1^{(1)} +y||\beta  h_2^{(1)} +y|}(1-\frac{ 2 |{\vartheta}^{(1)}|^2 e^{2u^{(1)}(t)} }{\sum_i |{\vartheta}^{(i)}|^2  e^{2u^{(i)}(t)}})\cos(\omega_{12}^{(11)}-\gamma^{(1)})\nonumber\\
p'_{12} &=& x+\frac{\alpha}{2}(h^{(1)}_{1}+ h^{(1)}_{2})-  \sqrt{|\beta  h_1^{(1)} +y||\beta  h_2^{(1)} +y|}(1-\frac{ 2 |{\vartheta}^{(1)}|^2 e^{2u^{(1)}(t)} }{\sum_i |{\vartheta}^{(i)}|^2  e^{2u^{(i)}(t)}})\sin(\omega_{12}^{(11)}-\gamma^{(1)})\nonumber\\
p_{13} &=& x +\frac{\alpha}{2}(h^{(1)}_{1}+ h^{(2)}_{1})+\frac{ 2 {\rm sgn}(\beta){\vartheta}^{(1)} \overline{{\vartheta}^{(2)}} e^{[u^{(1)}(t) + u^{(2)}(t)] }}{\sqrt{( h_1^{(1)} -  h_2^{(1)})( h_1^{(2)} -  h_2^{(2)})}\sum_i |{\vartheta}^{(i)}|^2  e^{2u^{(i)}(t)}}\nonumber\\ &\times&\sqrt{|\beta h_2^{(1)} + y||\beta h_2^{(2)} +y|}( h_1^{(2)} -  h_1^{(1)})\sin(\omega_{11}^{(12)}-[v^{(1)}(t) - v^{(2)}(t)]-(\gamma^{(1)} - \gamma^{(2)}))\nonumber\\
p'_{13} &=& x +\frac{\alpha}{2}(h^{(1)}_{1}+ h^{(2)}_{1})+\frac{ 2 {\rm sgn}(\beta){\vartheta}^{(1)} \overline{{\vartheta}^{(2)}} e^{[u^{(1)}(t) + u^{(2)}(t)] }}{\sqrt{( h_1^{(1)} -  h_2^{(1)})( h_1^{(2)} -  h_2^{(2)})}\sum_i |{\vartheta}^{(i)}|^2  e^{2u^{(i)}(t)}}\nonumber\\ &\times&\sqrt{|\beta h_2^{(1)} + y||\beta h_2^{(2)} +y|}( h_1^{(2)} -  h_1^{(1)})\cos(\omega_{11}^{(12)}-[v^{(1)}(t) - v^{(2)}(t)]-(\gamma^{(1)} - \gamma^{(2)}))\nonumber
\ee
The dynamics of all the 30 populations is illustrated in Fig.~\ref{Fig6} for a simple `seasonal' change of coupling of the species with their environment, $F(t)=\sin\omega t$. Fig.~\ref{Fig11} shows what happens if one takes a non-periodic  $F(t)=\sinh\omega t$. Of course, the solution is valid for any $F(t)$.

\subsection{Building seed solutions with $3\times 3$ blocks}

In one of our previous examples the seed solution for a $3\times 3$ $\rho$ was constructed by means of a $2\times 2$ block, trivially extended to $3\times 3$ by adding a column and row consisting of zeros. Here we explicitly construct a less trivial $3\times 3$ block. Of crucial importance is again the degeneracy condition, which means that the seed $\rho^{(k)}$ have to correspond to the same $\mu$ and $z_\mu$ for all $k$. Otherwise one would not be able to combine $\rho^{(k)}$ from different blocks, acting in different invariant subspaces of $H$. 

Let us start with a Hermitian $3\times 3$ matrix
 \be
 \rho &=&  \left(\begin
 {array}{lcr}
 \rho_1 &a &b\\
 \bar{a} &\rho_2 &c\\
 \bar{b} &\bar{c} &\rho_3
 \end {array}\right).\nonumber
 \ee
 The equation 
 \be
 i\dot{\rho}  = [H,\theta_2 \rho^2 + \theta_1 \rho ]\label{nvne}
 \ee
is equivalent to
  \be
 \dot\rho_1 &=& \dot{\rho_2} = \dot{\rho_3} = 0\nonumber,\\
 i\dot{a} &=&  (h_1-h_2)[(\rho_1 +\rho_2)\theta_2 + \theta_1]a + \theta_2(h_1-h_2)b\bar{c} \nonumber,\\
 i\dot{b} &=&  (h_1-h_3)[(\rho_1 +\rho_2)\theta_2 + \theta_1]b + \theta_2(h_1-h_3)ac \nonumber,\\
 i\dot{c} &=&  (h_2-h_3)[(\rho_1 +\rho_2)\theta_2 + \theta_1]c + \theta_2(h_2-h_3)\bar{a}b\nonumber,
 \ee
a dynamical system formally similar to Euler's equations from classical mechanics of a rigid body (the `Euler-Arnold top' \citep{Arnold}). Let us now find the dynamics of modules and phases  of the  matrix elements:
 \be
 |a|^2 \frac{d}{dt}{\arg(a)} &=&  -(h_1-h_2)[(\rho_1 +\rho_2)\theta_2 + \theta_1]|a|^2 - t_{23}{\rm Re}(a\bar{b}c)\nonumber,\\
 |b|^2 \frac{d}{dt}{\arg(b)} &=&  -(h_1-h_3)[(\rho_1 +\rho_3)\theta_2 + \theta_1]|b|^2 - t_{13}{\rm Re}(a\bar{b}c)\nonumber,\\
 |c|^2 \frac{d}{dt}{\arg(c)} &=&  -(h_2-h_3)[(\rho_2 +\rho_3)\theta_2 + \theta_1]|c|^2 - \theta_2(h_2-h_3){\rm Re}(a\bar{b}c),\nonumber\\
 \frac{d}{dt}{|a|^2} &=&  -2(h_1-h_2)\theta_2{\rm Im}(a\bar{b}c)\nonumber,\\
  \frac{d}{dt}{|b|^2} &=&  2(h_1-h_3)\theta_2{\rm Im}(a\bar{b}c)\nonumber,\\
  \frac{d}{dt}{|c|^2} &=& -2(h_2-h_3)\theta_2{\rm Im}(a\bar{b}c)\nonumber,
 \ee
 and real and imaginary parts of them,
\be
 \frac{d}{dt}{{\rm Re}(a\bar{b}c)} &=& [\rho_1(h_3-h_2) +\rho_2(h_1-h_3) + \rho_3(h_2-h_1)]\theta_2{\rm Im}(a\bar{b}c)\nonumber\\
 \frac{d}{dt}{{\rm Im}(a\bar{b}c)} &=& -[\rho_1(h_3-h_2) +\rho_2(h_1-h_3) + \rho_3(h_2-h_1)]\theta_2{\rm Re}(a\bar{b}c)\nonumber\\
 &\pp=& -\theta_2(h_1-h_2)|b|^2|c|^2 + \theta_2(h_1-h_3)|a|^2|c|^2 - \theta_2(h_2-h_3)|a|^2|b|^2\nonumber
 \ee
We can find integrals of motion  of (\ref{nvne}), other than the diagonal elements of $\rho$, using the fact that four of the equations have right-hand sides proportional to the same time-dependent function:
  \be
 \dot X &=& \frac{d}{dt}\left[(h_1-h_3)|a|^2 + (h_1-h_2)|b|^2\right] =  0\nonumber\\
 \dot Y &=& \frac{d}{dt}\left[(h_2-h_3)|a|^2 - (h_1-h_2)|c|^2\right] =  0\nonumber\\
 \dot Z &=& \frac{d}{dt}\left[(h_1-h_3)|c|^2 + (h_2-h_3)|b|^2\right] = 0\nonumber\\
 \dot Q &=& \frac{d}{dt}\left[|a|^2 + |b|^2 + |c|^2\right] = 0\nonumber\\
 \dot U &=& \frac{d}{dt}\left([\rho_1(h_3-h_2) +\rho_2(h_1-h_3) + \rho_3(h_2-h_1)]|a|^2 + 2(h_1-h_2){\rm Re}(a\bar{b}c)\right) =  0\nonumber\\
 \dot V &=& \frac{d}{dt}\left([\rho_1(h_3-h_2) +\rho_2(h_1-h_3) + \rho_3(h_2-h_1)]|b|^2 - 2(h_1-h_3){\rm Re}(a\bar{b}c))\right) =  0\nonumber\\
 \dot W &=& \frac{d}{dt}\left([\rho_1(h_3-h_2) +\rho_2(h_1-h_3) + \rho_3(h_2-h_1)]|c|^2 + 2(h_2-h_3){\rm Re}(a\bar{b}c))\right) = 0\nonumber\label{uvw}
 \ee
They are not linearly independent, since
\be
X-Y &=& (h_1-h_2)Q\nonumber,\\
X+Z &=& (h_1-h_3)Q\nonumber,\\
Y+Z &=& (h_2-h_3)Q\nonumber,\\
U+V+W &=& [\rho_1(h_3-h_2) +\rho_2(h_1-h_3) + \rho_3(h_2-h_1)]Q\nonumber.
\ee
The next integrals of motion can be found from the characteristic polynomial,
\be
\det(\rho - \lambda I)&=& -\lambda^3 +\lambda^2(\rho_1 +\rho_2+\rho_3)
\nonumber\\
 &\pp=&-\lambda(\rho_1\rho_2 +\rho_1\rho_3+\rho_2\rho_3 - |a|^2 - |b|^2 - |c|^2)\nonumber\\
 &\pp=&+
\rho_1\rho_2\rho_3 +2{\rm Re}(a\bar{b}c) -|a|^2\rho_3-|b|^2\rho_2-|c|^2\rho_1\nonumber.
\ee
From time invariance of spectrum of solutions of (\ref{nvne}) we find the constant of motion
\be
R&=& |a|^2\rho_3+|b|^2\rho_2+|c|^2\rho_1 - 2{\rm Re}(a\bar{b}c)= \frac{1}{h_1-h_2}(\rho_2X+\rho_1Y-U)\nonumber\\
&=&\frac{1}{h_1-h_3}(\rho_3X+\rho_1Z+V)=\frac{1}{h_2-h_3}(\rho_3Y+\rho_2Z-W)\nonumber.
\ee

Now we are in position to verify which solution can be used in the dressing transformation. The first equation (\ref{zs3}) of the Lax pair is an eigenvalue problem for the operator $\frac{1}{\mu}\rho +H$. We expect  that the operator has at least one eigenvalue which is constant in time. Again, we should examine the characteristic polynomial,
\be
&&\det\left(\frac{1}{\mu}\rho +H - \lambda I\right)\nonumber\\
&&\pp==
 -\lambda^3 +\lambda^2\left(\frac{1}{\mu}(\rho_1 +\rho_2+\rho_3)+h_1+h_2+h_3\right)\nonumber\\
&&\pp{==}-\lambda\Bigg(\left(\frac{\rho_1}{\mu}-h_1\right)\left(\frac{\rho_2}{\mu}-h_2\right) +\left(\frac{\rho_1}{\mu}-h_1\right)\left(\frac{\rho_3}{\mu}-h_3\right)
\nonumber\\
&&
\pp{==}+\left(\frac{\rho_2}{\mu}-h_2\right)\left(\frac{\rho_3}{\mu}-h_3\right) -\frac{1}{\mu^2}\big( |a|^2 + |b|^2 + |c|^2\big)\Bigg)\nonumber\\
&&\pp{==}+
\left(\frac{\rho_1}{\mu}-h_1\right)\left(\frac{\rho_2}{\mu}-h_2\right)\left(\frac{\rho_3}{\mu}-h_3\right)\rho_1 
\nonumber\\
&&\pp{==}
+\frac{2}{\mu^3}{\rm Re}(a\bar{b}c) -\frac{|a|^2}{\mu^2}\left(\frac{\rho_3}{\mu}-h_3\right)-\frac{|b|^2}{\mu^2}\left(\frac{\rho_2}{\mu}-h_2\right)-\frac{|c|^2}{\mu^2}\left(\frac{\rho_1}{\mu}-h_1\right)\nonumber
\ee
The coefficients of the polynomial are constant as one can easily verify. The last one can be described in terms of invariants,
\be
&{}&\left(\frac{\rho_1}{\mu}-h_1\right)\left(\frac{\rho_2}{\mu}-h_2\right)\left(\frac{\rho_3}{\mu}-h_3\right)\rho_1 
\nonumber\\
&\pp=&\pp=
+\frac{2}{\mu^3}{\rm Re}(a\bar{b}c) -\frac{|a|^2}{\mu^2}\left(\frac{\rho_3}{\mu}-h_3\right)-\frac{|b|^2}{\mu^2}\left(\frac{\rho_2}{\mu}-h_2\right)-\frac{|c|^2}{\mu^2}\left(\frac{\rho_1}{\mu}-h_1\right)
\nonumber\\
&{}&\pp{==} =\left(\frac{\rho_1}{\mu}-h_1\right)\left(\frac{\rho_2}{\mu}-h_2\right)\left(\frac{\rho_3}{\mu}-h_3\right)\rho_1 
 - \frac{1}{\mu^3}R -S ,\nonumber
\ee
where $S=|a|^2h_3-|b|^2h_2-|c|^2h_1 = h_1Q-X=h_2Q-Y=h_3 Q+Z$. Therefore, any solution of (\ref{nvne}) in dimension three has all the eigenvalues constant. This means that 
the evolution of eigenvectors  is given by some $V^{-1}(t)$, and thus $\frac{1}{\mu}\rho(t) +H = V(t)[\frac{1}{\mu}\rho(0) +H]V^{-1}(t)$.

To construct an explicit dressing transformation we need to find a relatively simple solution for the seed $\rho$. Assume that $\rho^2$ is affine in $\rho$, with the constant term commuting with $H$, 
\be
\rho^2 = A_1\rho + A_2(\rho H + H\rho ) + A_3 H + A_4 H^2 + A_5 I \label{r2}.
\ee
Then
\be
[H,\theta_2\rho^2+\theta_1\rho+ \theta_0 I] = [(\theta_1 + \theta_2 A_1)H + \theta_2 A_2 H^2, \rho]\nonumber,
\ee
so the equation for $\rho$, with $f(\rho)=\theta_2 \rho^2 + \theta_1 \rho + \theta_0 I$, becomes linear
\be
i\dot{\rho} &=&[H,f(\rho)]=[(\theta_1 + \theta_2 A_1)H + \theta_2 A_2 H^2, \rho]\nonumber.
\ee
Its solution reads
\be
\rho (t)&=& U(t)\rho(0)U^\dag (t)\nonumber
\ee
with
\be
U(t) = \exp\left(-i\left[\int_0^t(\theta_1 + \theta_2 A_1)d\tau H + \int_0^t\theta_2 A_2 d\tau H^2 \right] \right)=\exp\left(-i\int_0^tG d\tau  \right )\nonumber.
\ee
(\ref{r2}) is equivalent to the system of six equations
 \be
 \rho_1^2+ |a|^2 + |b|^2 &=&  A_1 \rho_1 +2A_2 \rho_1 h_1 +A_3 h_1 +A_4 h_1^2 + A_5\nonumber,\\
 \rho_2^2+ |a|^2 + |c|^2 &=& A_1 \rho_2 +2A_2 \rho_2 h_2 +A_3 h_2 +A_4 h_2^2 + A_5\nonumber,\\
 \rho_3^2+ |c|^2 + |b|^2 &=& A_1 \rho_3 +2A_2 \rho_3 h_3 +A_3 h_3 +A_4 h_3^2 + A_5\nonumber,\\
 (\rho_1+\rho_2)a + b\bar{c} &=& A_1 a + A_2 a(h_1+h_2)\label{r3},\\
 (\rho_1+\rho_3)b + ac &=& A_1 b + A_2 b(h_1+h_3) \label{r4},\\
 (\rho_2+\rho_3)c + \bar{a}b &=& A_1 c + A_2 c(h_2+h_3)\label{r5}.
 \ee
Let $a=|a|e^{i\varsigma_a}$, $b=|b|e^{i\varsigma_b}$ and $c=|c|e^{i\varsigma_c}$, then from (\ref{r3})-- (\ref{r5})
follow $b\bar{c}=Sa$, $ac =Sb$ and $\bar{a}b=Sc$ for some $S$. It is possible only for $|a|=|b|=|c|$ and $\varsigma_b= \varsigma_a + \varsigma_c$.
Eliminating $a$ from (\ref{r3}) we find that $\rho_j=p h_j+z$, for some parameters $p$ and $z$, so
\be
\rho &=&
\left(
 \begin{matrix}
 z +p h_1&|a|e^{i\varsigma_a}&|a|e^{i(\varsigma_a + \varsigma_c)}\\
 |a|e^{-i\varsigma_a} & z +p h_2&|a|e^{i\varsigma_c}\\
 |a|e^{-i(\varsigma_a + \varsigma_c)}&|a|e^{-i\varsigma_c}& z +p h_3
 \end{matrix}
 \right),\label{3x3blok}
 \ee
 and
 \be
 A_1 &=& 2z +|a|\nonumber,\\
 A_2 &=& p\nonumber,\\
 A_3 &=& -p(|a|+2z)\nonumber,\\
 A_4 &=& -p^2\nonumber,\\
 A_5 &=& 2|a|^2 -|a|z -z^2\nonumber.
 \ee
Accordingly, 
\be
\rho^2 &=& (2z +|a|)\rho + p(\rho H + H\rho ) -p(|a|+2z) H -p^2 H^2 + (2|a|^2 -|a|z -z^2)I,\nonumber\\
G &=& (\theta_1 + \theta_2 (2z +|a|))H + \theta_2 p H^2.\nonumber
\ee
We now use the fact that $G$ commutes with $H$, to make the ZS spectral problem  static (exactly like in the two dimensional case). We define 
$|\phi \rangle = {U}^\dag |\varphi \rangle = e^{i\int_0^t G d\tau  }|\varphi \rangle$, so that
\begin{eqnarray}
z_{\mu }|\varphi \rangle &=& \left(\frac{1}{\mu }\rho -  H \right)|\varphi \rangle,\cr
i{|\dot \varphi \rangle} &=& \frac{1}{\mu }f(\rho )|\varphi \rangle ,\nonumber
\end{eqnarray}
implies (for $\rho$ at $t=0$),
\begin{eqnarray}
z_{\mu }|\phi \rangle &=& \left(\frac{1}{\mu }{\rho }(0)-  H\right)|\phi \rangle,\cr
i{|\dot \phi\rangle} &=& \left( \frac{1}{\mu }f({\rho }(0)) - G \right)
|\phi \rangle\nonumber.
\end{eqnarray}

Following the strategy from the two dimensional case we check that the generator of the dynamics of $|\phi \rangle$ commutes with the operator of the first ZS problem:
\be
\left[\frac{\rho}{\mu}-H, \frac{f(\rho)}{\mu}-G\right]=0.
\ee
The latter means that both operators are some functions of one another. Since any function of a $3\times 3$ matrix is equivalent to a second-order polynomial, then
\be
\frac{f(\rho)}{\mu}-G=B_1 \left(\frac{\rho}{\mu}-H\right)^2 +B_2\left(\frac{\rho}{\mu}-H\right)+B_3 I\nonumber
\ee
we can compare coefficients on both sides, arriving at
    \be
  B_1 &=&\frac{\theta_2 p\mu}{p-\mu} \nonumber,\\
  B_2 &=&\theta_1 + \theta_2 (2z +|a|)\frac{\mu}{\mu-p}\nonumber,\\
  B_3 &=&\theta_2(2|a|^2 -|a|z -z^2)\frac{1}{\mu-p} + \frac{\theta_0}{\mu}\nonumber.
  \ee
Finally,
  \be
 \frac{f(\rho)}{\mu}-G&=& \frac{\theta_2 p\mu}{p-\mu}\left(\frac{\rho}{\mu}-H\right)^2 
 +
 \left(\theta_1 + \theta_2 (2z +|a|)\left(1+ \frac{p}{\mu-p}\right)\right)\left(\frac{\rho}{\mu}-H\right)\nonumber\\
 &\pp=& +\left(\theta_2(2|a|^2 -|a|z -z^2)\frac{1}{\mu}\left(1+ \frac{p}{\mu-p}\right) + \frac{\theta_0}{\mu}\right)I\nonumber
\ee
A very simple evolution of $|\phi \rangle$ is the next consequence of this relation,
\begin{eqnarray}
i{|\dot\phi\rangle} &=& \left(\frac{1}{\mu }f({\rho }(0)) - G \right)
|\phi \rangle\nonumber\\
&=&
\Bigg[ \frac{\theta_2 p\mu}{p-\mu}\left(\frac{\rho}{\mu}-H\right)^2 +\left(\theta_1 + \theta_2 (2z +|a|)\left(1+ \frac{p}{\mu-p}\right)\right)\left(\frac{\rho}{\mu}-H\right)\nonumber\\
 &\pp=& +\left(\theta_2(2|a|^2 -|a|z -z^2)\frac{1}{\mu}\left(1+ \frac{p}{\mu-p}\right) + \frac{\theta_0}{\mu}\right)I\nonumber\Bigg]|\phi\rangle\nonumber\\
 &=& \Bigg[\frac{\theta_2 p\mu}{p-\mu}z^2_{\mu}+\big(\theta_1 + \theta_2 (2z +|a|)\big)\left(1+ \frac{p}{\mu-p}\right)z_{\mu}
 \nonumber\\ 
 &\pp=& +\left(\theta_2(2|a|^2 -|a|z -z^2)\frac{1}{\mu}\left(1+ \frac{p}{\mu-p}\right) + \frac{\theta_0}{\mu}\right)\Bigg]|\phi\rangle\nonumber
\end{eqnarray}
Exactly like in the two-dimensional case this vector changes only by a time-depended factor. To calculate it we first have to find an eigenvalue of $\frac{\rho}{\mu}-H$:
 \be
\det(\rho-\mu H -\lambda I) &=&
\left|
 \begin{matrix}
 z +(p-\mu) h_1-\lambda&|a|e^{i\varsigma_a}&|a|e^{i(\varsigma_a + \varsigma_c)}\\
 |a|e^{-i\varsigma_a} & z +(p-\mu) h_2-\lambda&|a|e^{i\varsigma_c}\\
 |a|e^{-i(\varsigma_a + \varsigma_c)}&|a|e^{-i\varsigma_c}& z +(p-\mu) h_3-\lambda
 \end{matrix}
 \right|\nonumber\\
&=& \left|
 \begin{matrix}
 -\lambda'&|a|e^{i\varsigma_a}&|a|e^{i(\varsigma_a + \varsigma_c)}\\
 |a|e^{-i\varsigma_a} & sg_1-\lambda'&|a|e^{i\varsigma_c}\\
 |a|e^{-i(\varsigma_a + \varsigma_c)}&|a|e^{-i\varsigma_c}& sg_2-\lambda'
 \end{matrix}
 \right|\nonumber\\
 &=&-\lambda'^3+\lambda'^2 s(g_1+g_2)+\lambda'(3|a|^2-s^2 g_1g_2)+2|a|^3-|a|^2s(g_1+g_2),\nonumber
 \ee
 where $\lambda'=\lambda-z-(p-\mu) h_1$ and  $g_i= h_{i+1}-h_1$. There are three eigenvalues, but let us choose the parameters in a way guaranteeing that one of the eigenvalues equals $|a|$. This implies that $p=\alpha$, $|a|=\frac{|\beta|\sqrt{-g_1g_2}}{2}$. In this case $\mu z_{\mu} =
 \frac{|\beta|\sqrt{-g_1g_2}}{2}+z -i\beta h_1$, and
 \be
i{|\dot\phi\rangle} &=& \Bigg( \theta_2 \left(\mu z^2_{\mu}+ h_1|a| -i\beta h_1^2
  - \frac{2i|a|^2}{\beta} \right)+ \frac{\theta_1}{\mu}\big(|a|+z -i\beta h_1\big) + \frac{\theta_0}{\mu}\Bigg)|\phi\rangle\nonumber.
\ee
The corresponding eigenvector reads
 \be
 |\phi(0)\rangle 
  &=&
  \frac{1}{\sqrt{2 |h_3-h_2|}} \left(
 \begin{matrix}
 (i{\rm sgn}(\beta)\sqrt{|h_3-h_1|}+ \sqrt{|h_2-h_1|})e^{i(\varsigma_a + \varsigma_c)}\\
 i{\rm sgn}(\beta) \sqrt{|h_3-h_1|}e^{i\varsigma_c}\\
 \sqrt{|h_2-h_1|}
 \end{matrix}
 \right)
 \nonumber\\
 &\equiv& \left(\begin{array}{c}  \phi_1^{(k)} (0)   \cr
\phi_2^{(k)} (0) \cr
\phi_3^{(k)} (0) \cr
\end{array}\right)
=
 |\phi(0)^{(k)}\rangle 
 \nonumber
 \ee
 The index $k$ reminds us that, from the point of view of applications, we consider a solution associated with a $k$th block; the only parameter that is block-independent is here $\beta$ (the imaginary part of $\mu$).
In order to  get the explicit evolution of $|\varphi\rangle$ we first calculate the time-dependent factor,
 \be
u(t) +i v(t) 
  &=& \int_0^t\theta_2(\tau)d\tau   \left[\frac{2\alpha x y +\beta(y^2 -x^2)}{\alpha^2+\beta^2} -\beta h_1^2
  - \frac{2|a|^2}{\beta} \right ]\nonumber\\
   &\pp=&+i\int_0^t\theta_2(\tau)d\tau   \left[\frac{\alpha(y^2 -x^2) -2\beta x y}{\alpha^2+\beta^2}- h_1|a|  \right]\nonumber\\
 &\pp=&+\beta\frac{\int_0^t\theta_1(\tau)d\tau[(\frac{|\beta|\sqrt{-g_1g_2}}{2}+ z) -\alpha h_1] + \int_0^t\theta_0(\tau)d\tau }{\alpha^2+\beta^2} \nonumber\\
 &\pp=&-i\frac{\int_0^t\theta_1(\tau)d\tau[\alpha(\frac{|\beta|\sqrt{-g_1g_2}}{2}+ z) +\beta^2 h_1] + \int_0^t\theta_0(\tau)d\tau\alpha }{\alpha^2+\beta^2} \nonumber
\ee
and thus
\be
  |\phi(t)\rangle=\frac{e^{u(t) +i v(t)}}{\sqrt{2\beta^2 g_2(g_2-g_1)}} \left(
 \begin{matrix}
  (i\beta g_2+ |\beta|\sqrt{-g_1g_2})e^{i(\varsigma_a + \varsigma_c)}\\
 i\beta g_2e^{i\varsigma_c}\\
  |\beta|\sqrt{-g_1g_2}
 \end{matrix}
 \right)
 =|\phi(t)^{(k)}\rangle
 \nonumber.
 \ee
Finally, we use the form of the operator $G=\left(\theta_1 + \theta_2 (2z +\frac{|\beta|\sqrt{-g_1g_2}}{2})\right)H + \theta_2 \alpha H^2$ to get the explicit solution of the ZS Lax pair,
\be
|\varphi (t)\rangle
 &=& e^{-i\int_0^t G(\tau) d\tau  }e^{u(t) +i v(t)}|\phi (0)\rangle\nonumber
\\
&=& e^{-i\left[\left(\int_0^t \theta_1(\tau) d\tau + \int_0^t\theta_2(\tau) d\tau (2z +\frac{|\beta|\sqrt{-g_1g_2}}{2})\right)H + \alpha \int_0^t\theta_2(\tau) d\tau H^2 \right] }|\phi(t)\rangle\nonumber
\\
&=& |\varphi (t)^{(k)}\rangle\nonumber.
\ee
To sum up this stage, we have found both a seed $\rho$, which satisfies the von Neumann equation, and the vector $|\varphi(t)\rangle$ that solves the Lax pair. 
To make sure that $\rho[1]$ will be a density matrix we have to impose additional restrictions on parameters of the seed solution.
By Sylvester's positivity criterion,
\be
 &&z +\alpha h_1>0,\nonumber\\
&&(z +\alpha h_1)(z +\alpha h_2)>|a|^2\nonumber,\\
&& (z +\alpha h_1)(z +\alpha h_2)(z +\alpha h_3)+2|a|^3 -|a|^2(3z+\alpha(h_1+h_2+h_3))>0.\nonumber
\ee
For $z +\alpha h_i>|a|=\frac{|\beta|\sqrt{-g_1g_2}}{2}$ the three inequalities are satisfied. 

Now consider the case 
\be
H=\textrm{diag}(h_1,\dots,h_7)=\textrm{diag}\left(h_1^{(1)},h_2^{(1)},h_3^{(1)},h_1^{(2)},h_2^{(2)},h_1^{(3)},h_2^{(3)}\right),
\ee
and
\be
z=\frac{1}{7}\left[ \alpha\left(1 - \sum_{i=1}^7 h_i\right)   - 2 \beta \sqrt{(h_1 - h_2) (h_3 - h_1)} \right].
\ee
The dressed solution has the block form
\be
\rho[1] = \left(\begin{array}{ccc} \rho[1]^{(11)}  & \rho[1]^{(12)}  & \rho[1]^{(13)} \\
\rho[1]^{(21)}  & \rho[1]^{(22)}  & \rho[1]^{(23)} \\
\rho[1]^{(31)} &\rho[1]^{(32)}  & \rho[1]^{(33)} 
\end{array}\right)
=
\left(\begin{array}{ccc} 3\times 3  & 3\times 2  & 3\times 2 \\
2\times 3  & 2\times 2  & 2\times 2 \\
2\times 3 &2\times 2  & 2\times 2
\end{array}\right)
\nonumber,
\ee
where at the right-hand-side the dimensions of the blocks have been indicated.
The 2-dimensional blocks are constructed by means of (\ref{64})--(\ref{67}), with $w^{(2)}=w^{(3)}=0$, and
\be
c_{kl}(t) =  \frac{  {\vartheta}^{(k)} \overline{{\vartheta}^{(l)}} e^{[u^{(k)}(t) + u^{(l)}(t)] + i [v^{(k)}(t) - v^{(l)}(t)]} }{\sum_{i=1}^3 |{\vartheta}^{(i)}|^2  e^{2u^{(i)}(t)}}.
\ee
Note the sum from 1 to 3 in the denominator of $c_{kl}$. The 3-dimensional block on the diagonal reads
\begin{eqnarray}
\rho[1]^{(11)} =\left(\begin{array}{ccc} \alpha h_1^{(1)} +z  & s_{12}^{(1)}e^{i\varsigma^{(1)}_a }e^{-i\omega_{12}^{(11)}}& s_{13}^{(1)}e^{i(\varsigma^{(1)}_a+ \varsigma^{(1)}_c)}e^{-i\omega_{13}^{(11)}} \cr
 s_{21}^{(1)}e^{-i\varsigma^{(1)}_a }e^{-i\omega_{21}^{(11)}}& \alpha h_2^{(1)} +z& s_{23}^{(1)}e^{i\varsigma^{(1)}_c }e^{-i\omega_{23}^{(11)}}  \cr
s_{31}^{(1)}e^{-i(\varsigma^{(1)}_a+ \varsigma^{(1)}_c)}e^{-i\omega_{31}^{(11)}} & s_{32}^{(1)}e^{-i\varsigma^{(1)}_c }e^{-i\omega_{32}^{(11)}}  & \alpha h_3^{(1)} +z\cr
\end{array}\right).\label{rho[1]^11}
\end{eqnarray}
The following functions have been introduced in (\ref{rho[1]^11}):
\be
s_{12}^{(1)}&=&|\beta|\sqrt{| h_2^{(1)} -  h_1^{(1)}|| h_3^{(1)} -  h_1^{(1)}|}\left(\frac{1}{2}+\frac{ c_{11}( h_2^{(1)} -  h_1^{(1)}) }{| h_3^{(1)} -  h_2^{(1)}|}\right)\nonumber\\ &\pp=& +i\beta c_{11}( h_2^{(1)} -  h_1^{(1)})\frac{| h_3^{(1)} -  h_1^{(1)}|}{| h_3^{(1)} -  h_2^{(1)}|} = \overline{s_{21}^{(1)}}\nonumber\\
s_{13}^{(1)}&=&|\beta|\sqrt{| h_2^{(1)} -  h_1^{(1)}|| h_3^{(1)} -  h_1^{(1)}|}\left(\frac{1}{2}-\frac{ c_{11}( h_3^{(1)} -  h_1^{(1)}) }{| h_3^{(1)} -  h_2^{(1)}|}\right) \nonumber\\ &\pp=& +i\beta c_{11}( h_3^{(1)} -  h_1^{(1)})\frac{| h_2^{(1)} -  h_1^{(1)}|}{| h_3^{(1)} -  h_2^{(1)}|}= \overline{s_{31}^{(1)}}\nonumber\\
s_{23}^{(1)}&=&|\beta|\sqrt{| h_2^{(1)} -  h_1^{(1)}|| h_3^{(1)} -  h_1^{(1)}|}\left(\frac{1}{2}-\frac{ c_{11}( h_3^{(1)} -  h_2^{(1)}) }{| h_3^{(1)} -  h_2^{(1)}|}\right)= \overline{s_{32}^{(1)}}\nonumber
\ee
and
 \be
 \omega_{ij}^{(11)} &=&\left(\int_0^t \theta_1(\tau) d\tau + \int_0^t\theta_2(\tau) d\tau (2z +\frac{|\beta|}{2}\sqrt{| h_2^{(1)} -  h_1^{(1)}|| h_3^{(1)} -  h_1^{(1)}|})\right)(h^{(1)}_i-h^{(1)}_j) \nonumber\\
 &\pp=&
 +p  \int_0^t\theta_2(\tau) d\tau [(h^{(1)}_i)^2  -(h^{(1)}_j)^2].\nonumber
\ee
The two $3\times 2$ blocks read
\begin{eqnarray}
\rho[1])^{(kl)} = \frac{ 2i \beta c_{kl}}{\sqrt{2|\beta|| h_3^{(k)} -  h_2^{(k)}|| h_1^{(l)} -  h_2^{(l)}|}}\left(\begin{array}{cc} r^{(kl)}_{11}e^{i(\varsigma^{(k)}_a + \varsigma^{(k)}_c - \gamma^{(l)})}  &
r^{(kl)}_{12} e^{i(\varsigma^{(k)}_a + \varsigma^{(k)}_c )} \cr
r^{(kl)}_{21}e^{i( \varsigma^{(k)}_c - \gamma^{(l)})} & r^{(kl)}_{22} e^{i \varsigma^{(k)}_c } \cr
r^{(kl)}_{31}e^{-i\gamma^{(l)} } & r^{(kl)}_{32} \cr
\end{array}\right)\nonumber,
\end{eqnarray}
where 
\be
r^{(kl)}_{ij} &=&  \hat{\phi}_i^{(k)}  \sqrt{|\beta h_{\sigma(j)}^{(l)} +y|}( h_j^{(l)} -  h_i^{(k)})e^{-i\omega_{ij}^{(kl)}},\nonumber\\
\hat{\phi}_1^{(k)} &=&
\left(i{\rm sgn}(\beta)\sqrt{|h_3-h_1|}+ \sqrt{|h_2-h_1|}\right)e^{i(\varsigma_a + \varsigma_c)},\nonumber\\
\hat{\phi}_2^{(k)} &=&  i{\rm sgn}(\beta) \sqrt{|h_3-h_1|}e^{i\varsigma_c},\nonumber\\
\hat{\phi}_3^{(k)} &=& \sqrt{|h_2-h_1|},\nonumber
\ee
($\sigma(i)$ is an odd permutation of $i\in \{ 1,2\}$), and
\be
\omega_{ij}^{(kl)} &=&\left(\int_0^t \theta_1^{(k)}(\tau) d\tau + \int_0^t\theta_2^{(k)}(\tau) d\tau \left(2z +\frac{|\beta|}{2}\sqrt{| h_2^{(k)} -  h_1^{(k)}|| h_3^{(k)} -  h_1^{(k)}|}\right)\right)h^{(k)}_i \nonumber\\
  &\pp=&+\alpha  \int_0^t\theta_2^{(k)}(\tau) d\tau (h^{(k)}_i)^2  - \int_0^t \theta_1^{(l)}(\tau)h^{(l)}_j d\tau\nonumber.
\ee
The remaining two $2\times 3$ blocks are the Hermitian conjugates of the $3\times 2$ ones, so that $\rho[1]$ is Hermitian. 

\begin{figure}
\centering
\includegraphics[width=.5\textwidth]{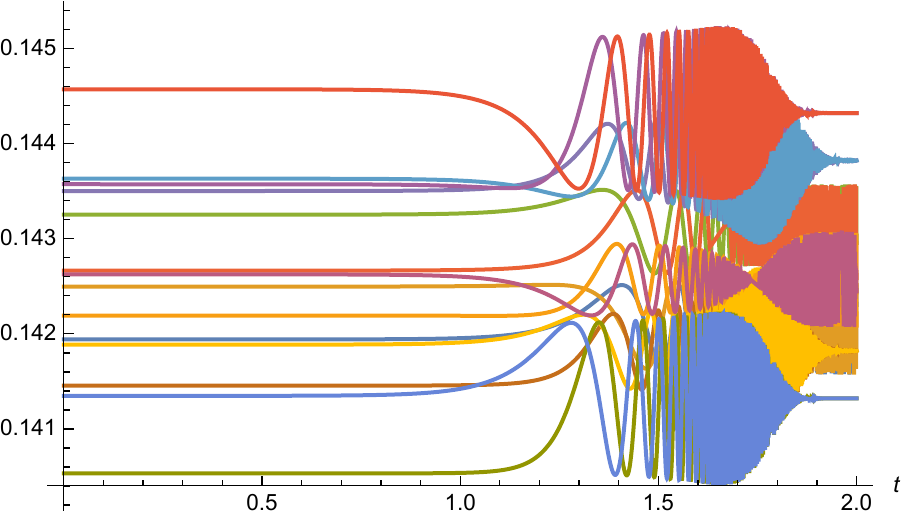}
\includegraphics[width=.5\textwidth]{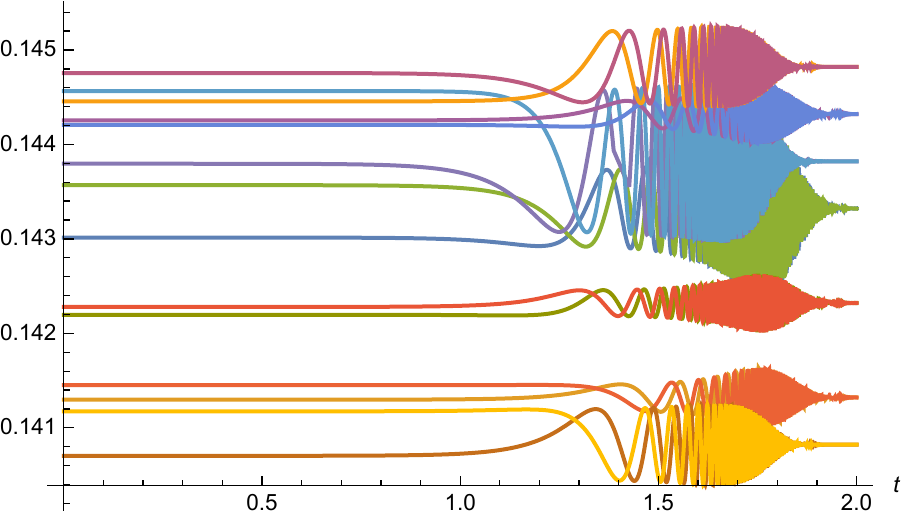}
\includegraphics[width=.5\textwidth]{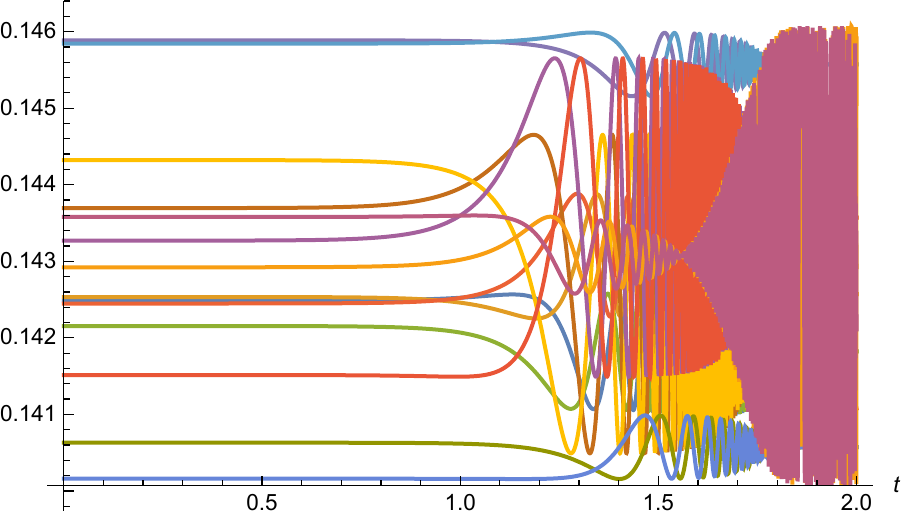}
\caption{42 populations for $f(\rho) = \sinh(\omega t) \rho^2$, $h_1=0.0051$, $h_2=0.005$, $h_3=0.0052$, $h_4=0.0053$, $h_5=0.0049$, $h_6=0.0054$, $h_7=0.0048$, $\vartheta^{(1)}=1+i$, $\vartheta^{(2)}=2+i$, $\vartheta^{(3)}=3+i$, $\alpha=10$, $\beta=10$, $\omega=10$, $\gamma_1=2$, $\gamma_2=3$, $\gamma_3=4$, $\gamma_4=5$ .}
\label{Fig77}
\end{figure}

Fig.~\ref{Fig77} shows the dynamics of all the 42 time-dependent populations associated with a $7\times 7$ solution $\rho[1]$. The diagonal elements are time-independent, so we do not plot them.

\section{Further perspectives}

We have developed a method of generating appropriate seed solutions for the soliton system $i\dot\rho(t)=\big[H,f_t\big(\rho(t)\big)\big]$. $H$ is Hermitian and its spectrum possesses a discrete part, an assumption valid for a large class of $H$. Differences of eigenvalues of $H$, $k_{jk}=h_j-h_k$, enter the rate equations as kinetic constants. So, given $k_{jk}$, one can design the dynamics by adjusting $H$. The nonlinearity $f_t$ is essentially arbitrary as well, although in practical computations one replaces $f_t$ by a polynomial with time dependent coefficients. From the point of view of realistic modeling the formalism is much more flexible than anything discussed in the literature so far. 

Still, is it flexible enough?  Probably not. Our equation is just a tip of the iceberg. Almost unexplored is  the general family of soliton von Neumann equations $i\dot\rho(t)=\big[H_t\big(\rho(t)\big),\rho(t)\big]$ discussed in \citep{CCU}. Here $H(\rho)$ is a `non-Abelian' function, a kind of polynomial whose coefficients are not numbers but operators. The relevant dressing transformations have been constructed  \citep{CCU}, but virtually nothing is known about the structure of seed solutions.

For time independent $f$ the system described by the von Neumann equation is conservative. A time dependent $f_t$ makes the system open: dissipative or driven. Dissipation can be introduced also through a time-independent part, such as the simple example discussed in \citep{ACKS},
\be
\dot\rho(t)=-i\big[H,f\big(\rho(t)\big)\big]+\xi \rho,
\ee
where $\xi\in\mathbb{R}$ is a birth or mortality rate. More generally, one can consider 
\be
\dot\rho(t)=-i\big[H\big(\rho(t)\big),\rho(t)\big]+\Xi(\rho),
\ee
where $\Xi(\rho)=\Xi(\rho)^\dag$ is a Hermitian, $\rho$-dependent operator, but little is known about soliton integrability of equations involving a less trivial 
$\Xi(\rho)$.

Yet another possibility of introducing dissipation without explicitly time-dependent parameters is to consider solutions analytically continued either in $t$ (a `complex time'), or in other parameters of the system. A complex time is known to turn the Schroedinger equation into a diffusion equation, so a similar trick can be performed in the von Neumann case \citep{Aerts Czachor(2007)}. But since Hermiticity of $\rho$ will be in general lost, a new model of probability has to be employed. 

\section*{Acknowledgment} 

I'm indebted to Marek Czachor for a critical reading of the entire manuscript. Special thanks to Diederik Aerts for various help during my stay in Centrum Leo Apostel in Brussels, where a part of this work was performed. 

\section*{Appendix: Replacing $f(\rho)$ by a polynomial}

Consider some Hermitian matrix $\rho$, or a Hermitian operator $\rho$ which has a finite number $n$ of different nonzero eigenvalues $\lambda_1, \dots, \lambda_n$. By spectral theorem $\rho=\sum_{j=1}^n \lambda_j P_j$, where $P_j$ are spectral projectors, and
$f(\rho)=\sum_{j=1}^n f(\lambda_j) P_j$. Now consider an arbitrary polynomial, 
\be
g(\rho) &=&\sum_{k=0}^K g_k\rho^k=\sum_{j=1}^n g(\lambda_j) P_j=\sum_{j=1}^n \sum_{k=0}^K g_k\lambda_j^k P_j.
\ee
In order to find a polynomial satisfying $f(\rho)=g(\rho)$ we have to solve for $g_k$ the system of linear equations
\be
f(\lambda_j) &=& \sum_{k=0}^K g_k\lambda_j^k, \quad j=1,\dots ,n,
\ee
with $K=n-1$.

\end{document}